\documentclass{nature}
\usepackage{pdfpages}
\usepackage{scicite}
\usepackage{amsmath}
\usepackage{graphicx}
\usepackage{xfrac}
\usepackage{multirow}
\usepackage{subfigure}
\usepackage{mathtools}
\usepackage[colorlinks=true,linkcolor=blue,citecolor=black,urlcolor=blue]{hyperref}
\usepackage{soul}
\usepackage{authblk}
\usepackage[labelfont=bf]{caption}
\usepackage[figurename=Figure ]{caption}
\usepackage{siunitx}
\usepackage{scalefnt}

\usepackage{lineno}

\makeatletter

\let\saved@includegraphics\includegraphics
\AtBeginDocument{\let\includegraphics\saved@includegraphics}
\renewenvironment*{figure}{\@float{figure}}{\end@float}
\makeatother

\usepackage{color}
\usepackage{dsfont}
\usepackage[normalem]{ulem}
\usepackage{dcolumn}
\usepackage{blindtext}
\usepackage{outlines}
\usepackage{enumitem}
\usepackage{soul}
\setlist[enumerate,2]{label=\roman*)}
\setlist[enumerate,3]{label=\alph*)}
\usepackage{multirow}
\newcommand{\MC}[1]{\mathcal{#1}}

\newcolumntype{d}[1]{D{.}{.}{#1}}
\definecolor{orange}{rgb}{1,0.5,0}
\definecolor{darkgreen}{RGB}{0,100,0}
\DeclareSIUnit\mub{\mu_\text{B}}
\DeclareSIUnit\rydberg{\text{Ry}}

\newcommand{\onlinecite}[1]{\hspace{-1 ex} \nocite{#1}\citenum{#1}}

\setstcolor{red}

\begin{document}

\title{Anomalous excitations of atomically crafted quantum magnets}

\author{Sascha Brinker$^{1,\ast}$, Felix K{\"u}ster$^{2}$,  Stuart S. P. Parkin$^{2}$, Paolo Sessi$^{2}$, Samir Lounis$^{1,3,\ast}$}
\affil{
$^1$Peter Gr{\"u}nberg Institut and Institute for Advanced Simulation, Forschungszentrum J{\"u}lich \& JARA, J{\"u}lich D-52425, Germany\\
$^2$Max Planck Institute of Microstructure Physics, Halle 06120, Germany\\
$^3$Faculty of Physics, University of Duisburg-Essen and CENIDE, 47053 Duisburg, Germany}

\affil{$^\ast$ Emails: s.brinker@fz-juelich.de,  s.lounis@fz-juelich.de }

\date{}

\maketitle


\begin{abstract}
High energy resolution spectroscopic studies of quantum magnets proved extremely valuable in accessing magnetodynamics quantities, such as energy barriers, magnetic interactions and lifetime of excited states.  
Here, we explore the evolution of a new flavor of low-energy spin-excitations for quantum spins coupled to an electron bath. In sharp contrast to the usual tunneling signature of two steps symmetrically centered around the Fermi level, we find a single step in the conductance. Combining time-dependent and many-body perturbation theories, magnetic field-dependent tunneling spectra are explained as  the result of an interplay between weak magnetic anisotropy energy, magnetic interactions and Stoner-like electron-hole excitations  that are strongly dependent on the magnetic states of the nanostructures. The results are rationalized in terms of a non-collinear magnetic ground state and the dominance of ferro- and antiferromagnetic interactions. The  atomically crafted nanomagnets offer an appealing model 
for the exploration of electrically pumped spin systems.
\end{abstract}

\subsection{Teaser:} Novel paradigmatic signature of spin-excitations offers a new exploration platform for electrically pumped spin systems.

\section*{Introduction}
Atomic-scale magnetodynamics is  at the cornerstone of spin-based nanoscale devices for future information technology. It defines how and on which timescales magnetic states can be controllably manipulated.  
The interaction of local spins with the local environment plays a crucial role in determining their properties. Orbital hybridization effects, charge transfer, and the presence of nearby impurities have all been reported to strongly impact the magnetic ground state, ultimately determining a range of magnetodynamic quantities including magnetic anisotropy, spin lifetime, and spin relaxation mechanisms.\cite{SGC1993,Loth:2012,Khajetoorians:2013b,Rau988,HBP2013,Lounis:2010,Khajetoorians:2011,Paul:2017,Ternes:2015,Donati318,Baumann:2015,Choi:2019,SPB2020}. The development of experimental techniques capable of directly capturing these properties in reduced dimensions allows one to clearly visualize the breakdown of  classical or semiclassical descriptions of magnetic phenomena at subnanometer scales, revealing the emergence of exquisite quantum mechanical effects.~\cite{Madhavan:1998,Hirjibehedin:2006,Nadj-Perge:2014,Spinelli:2014,Toscovic:2016,Wernsdorfer133} These achievements enabled the understanding of the ultimate limits of classical computational schemes while simultaneously setting the ground for testing ideas and concepts which might find direct application in innovative spin-based quantum computational schemes.~\cite{LGC2007,Khajetoorians:2011b,Loth:2012,Hermenau:2017,Willkeeaaq1543}.

In this context, scanning tunneling spectroscopy techniques (STS) occupy a primary role. They were 
particularly instrumental in detecting not only several ground-state phenomena but also a wealth of fundamental excitations such as those with magnetic~\cite{Heinrich:2004,Hirjibehedin:2006,Rossier:2009,Lounis:2010,Khajetoorians:2011,Chilian:2011,Khajetoorians:2013,Bryant:2013,Oberg:2014,Ternes:2015},  vibrational~\cite{Stipe1998}, polaronic~\cite{Pohlit:2018}, excitonic~\cite{Pommier:2019} or Kondo characteristics~\cite{Kondo:1964,hewson_1993,Madhavan:1998,Knorr:2002ig,Heinrich:2004,Ternes:2008}. 
The analysis of the distinct spectral shapes identified in tunneling transport sheds light on their origin and ideally a link to a specific excitation process. For instance, measurements in an applied magnetic field have proven extremely useful in disentangling magnetic and non-magnetic excitations. 
However, dissimilar and in some cases competing processes can be simultaneously at play, which can make the analysis complicated. 
This requires a careful theoretical-experimental investigation to unveil the fundamental mechanisms at play.

The Kondo effect versus spin-excitation paradigm is a particularly intriguing case.~\cite{Otte:2008,Chen:2009,Sarkozy:2009,Bouaziz:2020a} 
Both phenomena share a similar magnetic origin. However, they are profoundly different in nature, ultimately leading to opposing technological implications. 
On the one hand, spin-excitations emerge from the promotion of a magnetic moment from its ground state to various excited states. 
Since spin-excitations play a crucial role in how magnetic bits are written, understanding these processes is key in information technology. 
In tunneling experiments, inelastic spin-excitations are expected to appear in the differential conductance as symmetric steps centred around the Fermi energy, which shift in energy in response to an applied magnetic field.~\cite{Heinrich:2004} 
Their spectral shape encodes information directly related to the reading and manipulation of spins, such as the lifetime of the excited states and the related energy barriers (e.g. the magnetic anisotropy energy).~\cite{Hirjibehedin:2006,Lounis:2010,Khajetoorians:2011,Rau988}
On the other hand, the Kondo effect emerges from the entanglement of the spin of an atom with that of the surrounding bath of electrons~\cite{Kondo:1964,hewson_1993}, which leads to many-body screening of the spin below the energy scale defined by the Kondo temperature. 
Consequently, although of great fundamental interest in the context of many-body physics, the Kondo effect is of limited technological importance.
Spectroscopically, the Kondo effect is revealed in complicated spectral shapes that are generally analyzed as Fano resonances.~\cite{Madhavan:1998,Knorr:2002ig,Ternes:2008} 
Consequently, a clear identification of the underlying mechanisms is often complex since many fitting parameters need to be used, especially when multi-orbitals are involved.~\cite{surer:2012,dang:2016} 
There are several examples, where the same spectral shape is found whether one invokes a Kondo effect or simply considers the molecular/atomic orbitals or a combination of both, which sometimes can be arbitrary \cite{Nagaoka:2002,Franke940,KKS2014,Bouaziz:2020a}. 
In this context, recent ab-initio simulations challenged the Kondo versus conventional spin-excitation picture. By including spin-orbit coupling, the well-known zero-bias anomalies experimentally observed for Co adatoms on Cu, Ag and Au surfaces~\cite{Bouaziz:2020a} were systematically explained as gapped unconventional spin excitations, calling for new experiments to verify their origin.

In this contribution, we report on the emergence of anomalous spectroscopic signatures centered around the Fermi energy for Cr single atoms coupled to metallic niobium, their magnetic field dependence, and their evolution once adatoms are coupled through the creation of atomically crafted antiferromagnetic and ferromagnetic nanostructures.   Although their spectral shape might be reminiscent of Kondo-resonances, our first-principles simulations identify them as new paradigmatic spectroscopic manifestations of spin excitations. While the observation of spin-excitations of antiferromagnetic multi-spin nanostructures has so far  been explored to a great extent on  thin insulating substrates (see e.g. Refs.~\onlinecite{Hirjibehedin:2006,Spinelli:2014,Toscovic:2016}), here we demonstrate on a metallic substrate the design of antiferromagnetic or ferromagnetic magnets  composed of a few exchange-coupled Cr atoms  that enables the tailoring of the shape and linewidth of the corresponding spectroscopic features, which are intimately linked to the adatoms' weak magnetic anisotropy energy and the strong dependence of the excitation lifetime on the underlying magnetic texture.

\section*{Results}

    \subsection{Adatoms.}
    
        \begin{figure}
        	\centering
        	\includegraphics[width=\textwidth]{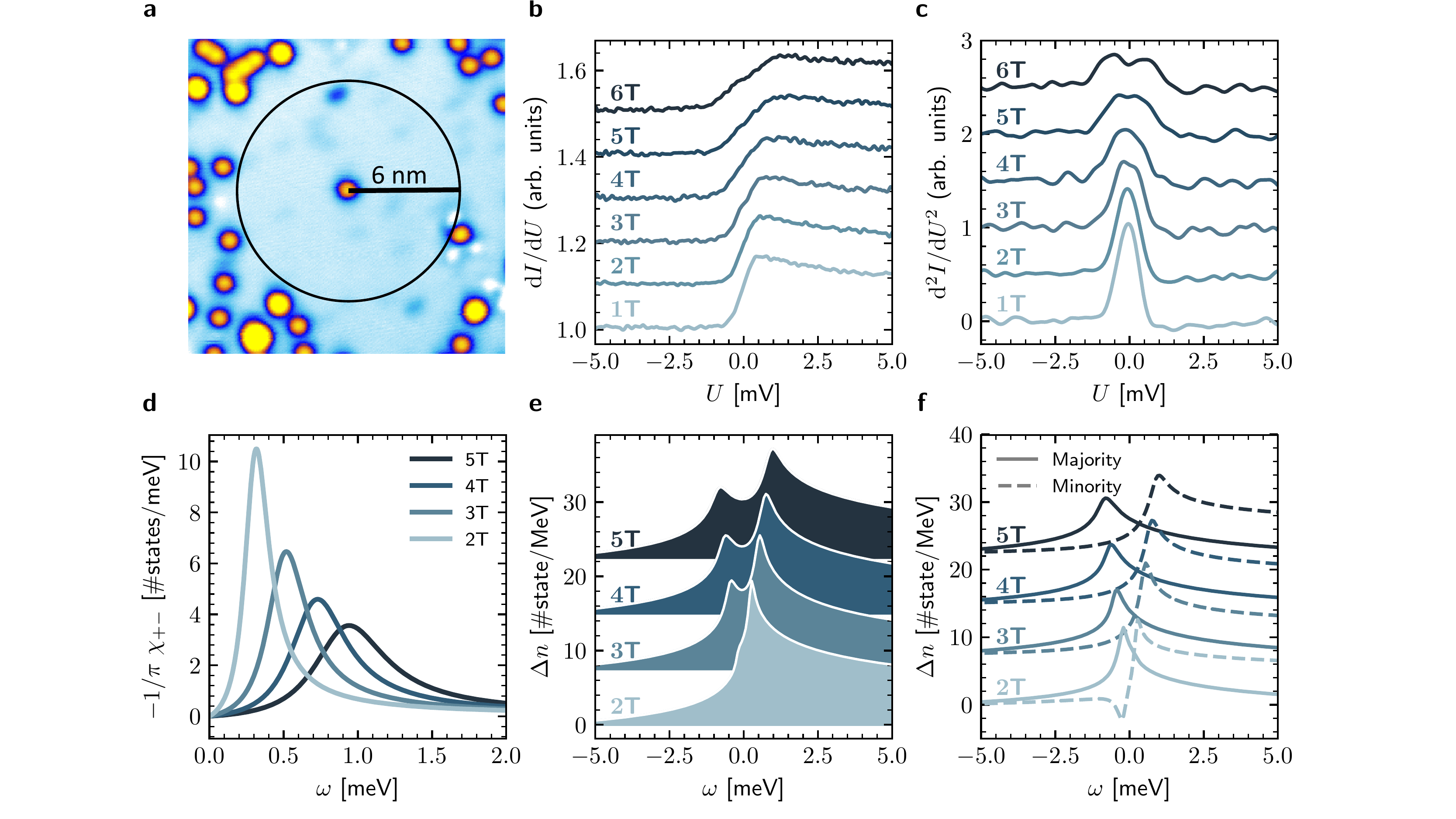}
        	\caption{
        	\textbf{Experimental and theoretical inelastic STS fingerprint of a Cr adatom on the Nb(110) surface with varying external magnetic fields.}
        	\textbf{a} Experimental topography image of the adatom.
        	\textbf{b} / \textbf{c} Experimental $\mathrm{d}I/\mathrm{d}U$ and $\mathrm{d}^2I/\mathrm{d}U^2$ spectra, $I=\SI{3}{\nano\ampere}$ and $T=\SI{600}{\milli\kelvin}$.
        	\textbf{d} Density of spin-flip excitations obtained from TD-DFT.
        	\textbf{e} Theoretical inelastic STS spectrum corresponding to the renormalized local density of states from MBPT calculated at a height of \SI{4.7}{\angstrom} above the adatom.
        	\textbf{f} Similar to \textbf{e} with the spin-resolved theoretical spectra.
        	The different field-dependent spectra are shifted to improve visibility.
        	} \label{fig:adatoms}
        \end{figure}
        
        As a prototypical platform, we focus on Cr atoms coupled to the (110) surface of Nb.  
        The Nb crystal was prepared according to the procedures described in Ref.~\onlinecite{Odobesko2019}.
        Single Cr adatoms were deposited directly onto a cold substrate, as described in the Methods section. 
        All atoms are found to be adsorbed in hollow stacking sites, which are energetically most favourable~\cite{Kuester2021}.
        Figure ~\ref{fig:adatoms}a shows a region of the Nb surface where blue dots correspond to Cr adatoms. 
        After deposition, atomic manipulation has been used to create a nanostructure with an isolated Cr atom, which is well-separated (distance $> \SI{6}{\nano\meter}$) from all surrounding atoms as shown in Fig.~\ref{fig:adatoms}a. 
        This allows us to analyze the effect of local magnetic moments coupled to an electron bath, while excluding the influence of nearby adatoms which can couple to each other by indirect interactions mediated by the conduction electrons, i.e. the RKKY coupling, ultimately influencing the magnetic adatom ground state \cite{kuster2021long}.
        
        Figure~\ref{fig:adatoms}b shows a sequence of spectra, measured in a small energy range centered around the Fermi level, where the tip was positioned directly on top of the Cr adatom. 
        All spectroscopic data have been acquired in the presence of a magnetic field applied perpendicular to the sample surface. 
        Starting from a value of \SI{1}{\tesla}, which is necessary to drive the otherwise superconducting Nb substrate ($T_\mathrm{c}=\SI{9.3}{\kelvin}$ and $B_\mathrm{c} = \SI{0.8}{\tesla}$) into the normal metallic regime, the strength of the magnetic field is progressively increased up to \SI{6}{\tesla}.  
        This procedure allows us to map the evolution of the spectroscopic data and to clearly assess the magnetic origin of the excitations. 
        The $\mathrm{d}I/\mathrm{d}U$ spectra are characterized by an asymmetric step-like feature around the Fermi level which shows a significant magnetic field-dependence. 
        The step starts at negative bias voltage and reaches its maximum at positive bias. 
        By increasing the magnetic field, the step broadens. 
        Moreover, an additional feature emerges at negative voltage starting from \SI{3}{\tesla}. 
        This effect is more clearly seen by looking at the derivative ($\mathrm{d}^2 I/\mathrm{d}U^2$) reported in Fig.~\ref{fig:adatoms}c, where the magnetic field-dependent broadening is clearly visible. At the maximum magnetic field available in our set-up, the broadening appears to split into two distinct features symmetrically centered around the Fermi level (additional experimental plots and fits are presented in Supplementary Notes 2--4). 
        As described in the following, spin-excitations can account for both the spectral shape as well as its magnetic field dependence. 

        To explore the origin of the anomalous spectral feature, we use a combination of relativistic time-dependent density functional theory (TD-DFT) and many-body perturbation theory (MBPT) implemented in the framework of the Korringa-Kohn-Rostoker Green function method~\cite{Schweflinghaus:2014,Bouaziz:2020a,Bouaziz:2020} (see Methods section). The ab-initio nature of the theoretical approach allows us to address the investigated spectroscopic phenomena without the use of adjusting parameters. Thus, we do not perform or rely on a fitting procedure of the experimental data. 
        This framework is ideal to treat the ground state and spin-excited state of adatoms on surfaces as shown in various studies~\cite{Lounis:2010,Khajetoorians:2011,Khajetoorians:2013} as well as their theoretical inelastic tunneling spectra~\cite{Schweflinghaus:2014,Schweflinghaus:2016,Bouaziz:2020a}. 
        The Cr adatom carries a spin moment of \SI{3.53}{\mub} characterized by a weak magnetic anisotropy energy, \SI{0.20}{\milli\electronvolt}, slightly favoring an in-plane orientation of the magnetic moment (the local density of states is presented in Supplementary Note 1). 
        However, in the presence of a magnetic field perpendicular to the substrate, the spin moment points out-of-plane. 
        Owing to the weak magnetic anisotropy of the Cr spin moment, it can be excited by the tunneling electrons at very low energies. 
        Our theory allows us to effectively capture these processes by quantifying the excitation probability in terms of the density of spin-flip excitations, which can be extracted from the imaginary part of the transverse spin-spin susceptibility, $-1/\pi \, \mathrm{Im} \, \chi_{+-}(\omega)$, shown in Fig.~\ref{fig:adatoms}d.
        The density of spin-excitations encodes two main features: (i) the energy of the spin excitation $\omega_0$ directly related to the magnetic anisotropy energy and applied magnetic field, and (ii) their lifetime, which is inversely proportional to the broadening (full width half maximum). 
        The origin of this broadening are Stoner-like electron-hole excitations provided by the highly itinerant electrons of the metallic substrate. Such single particle excitations are emitted while strongly contributing to the decay of spin-excitations.~\cite{Mills:1967,Lounis:2010}. 
        
        In contrast to a scenario where the adatoms are deposited on a insulator, the magnetic moments on a metal experience additional effects with respect to usual interactions forming a Heisenberg Hamiltonian such as the magnetic anisotropy, Zeeman energy and magnetic exchange interactions. The reason is the nature of electrons forming moments, which are delocalized and extend beyond the atomic site, and thus do not respond uniformly to external perturbation. Therefore, while the magnetic moment undergoes a spin-flip process, which is by nature a collective excitation process involving many electrons, single particle excitations, wherein a single electron may be excited from a majority-spin state below the Fermi level to a minority-spin state above, kick in and impact the overall dynamical behavior of the magnetic moment.  The inhomogeneous reaction of the electrons broaden then the ideal response of the spin-moment, leading to an effective decay, or damping, of the spin-excitation. 
        We note that such electron-hole excitations are associated with spin-currents discussed in the context of spin-pumping experiments~\cite{Tserkovnyak:2002,Zhang2015,Rojas:2014}.  The larger the number of electron-hole excitations, the stronger the damping and the shorter is the lifetime of the spin-excited states. 
        As  can be recognized from Fig.~\ref{fig:adatoms}d, our model effectively accounts for all experimental trends, both in the energy shift as well as in the broadening (Fig.~\ref{fig:adatoms}b-c). 
        
        When the magnetic moment of the atom is excited, the triggered spin-excitations affect the electron density resulting in the spectroscopic signatures visible in the theoretically calculated tunneling spectra reported in Fig.~\ref{fig:adatoms}e, which are obtained from a many-body perturbation theory (see Methods).
        In good agreement with the experimental measurements, we recover a step-like signature of the spin excitation  with a clear field-dependent broadening which progressively evolves towards a double peak structure at high magnetic field.   
        The origin of this behavior is effectively captured in Fig.~\ref{fig:adatoms}f, which illustrates the magnetic field evolution of the spin-resolved tunneling spectra for majority (continuous line) and minority (dashed line) spins. 
        A single peak feature at negative energies and a step-like function with onset at positive energies are visible for majority and minority spin channels, respectively. 
        The energy-separation between the majority and minority spectral signatures becomes larger by progressively increasing the strength of the magnetic field. We note that besides Cr adatom, Mn also displays a similar but weaker  step-like  zero-bias anomaly, which is nicely reproduced by our ab-intio simulations as shown in Supplementary Note 5.
        
        To grasp analytically the connection between the various magnetodynamical parameters shaping the unveiled spectroscopic anomalies,  it is useful to map 
        the TD-DFT results to the equation of motion of a spin-moment, the Landau-Lifshitz-Gilbert equation~\cite{Landau1935,Gilbert2004,Manuel:2015}, $\frac{d\mathbf{M}}{dt} =  \gamma \mathbf{M}\times\frac{d{H}}{d\mathbf{M}} + \frac{\alpha}{M} \mathbf{M}\times\frac{d\mathbf{M}}{dt}$  subjected to a Heisenberg hamiltonian $H$ (more details are given in the Method section and Supplementary Notes 6 and 8). 
        The field dependence of the excitation energy obeys $\omega_0 = \gamma'  (\MC{K} + BM) /M $, while the broadening (inverse lifetime) 
        of the excitation is given by $\alpha \gamma' (\MC{K} + BM)/M$, where $\MC{K}$ is the magnetic anisotropy energy which represents the costs in energy in rotating the magnetic moments from $z$ towards the $x$ and $y$ directions (corresponding to the $[1\overline{1}0]$ and $[001]$ direction, respectively). 
        $M$ is the magnetic moment, $B$ is an out-of-plane magnetic field, $\gamma'$ is the renormalized gyromagnetic ratio, and $\alpha$ is the so-called Gilbert damping parameter, which quantifies the probability of creating electron-hole excitations responsible for the decay of the spin-excitations. 
        This theoretical description allows us to effectively capture the following trends:  
        (i) the excitation energy becomes higher by increasing the strength of the magnetic field acting on the magnetic moment, (ii) this is accompanied by the damping that becomes progressively stronger by enhancing the amplitude of the electron-hole excitations. 
        Overall, this approach allows the visualization of the existence of competing trends in the manifestation of spin-excitations. 
        While an enhanced magnetic anisotropy facilitates their detection, it also simultaneously enables the creation of more electron-hole pairs, increasing the broadening, ultimately washing out their spectroscopic signature. 
        Overall, this could explain why identifying the spin-excitation origin of similar spectra observed for Co on Cu(001) and Ag(001) surfaces~\cite{Knorr:2002ig,Wahl:2004jy,Bouaziz:2020a}, characterized by a larger magnetic anisotropy energy than for Cr on Nb(110), is challenging since this requires extremely large magnetic fields and therefore has escaped experimental observation to date.

    \subsection{Dimers.}
        \begin{figure}[!ptbh]
        	\centering
        	\includegraphics[width=\textwidth]{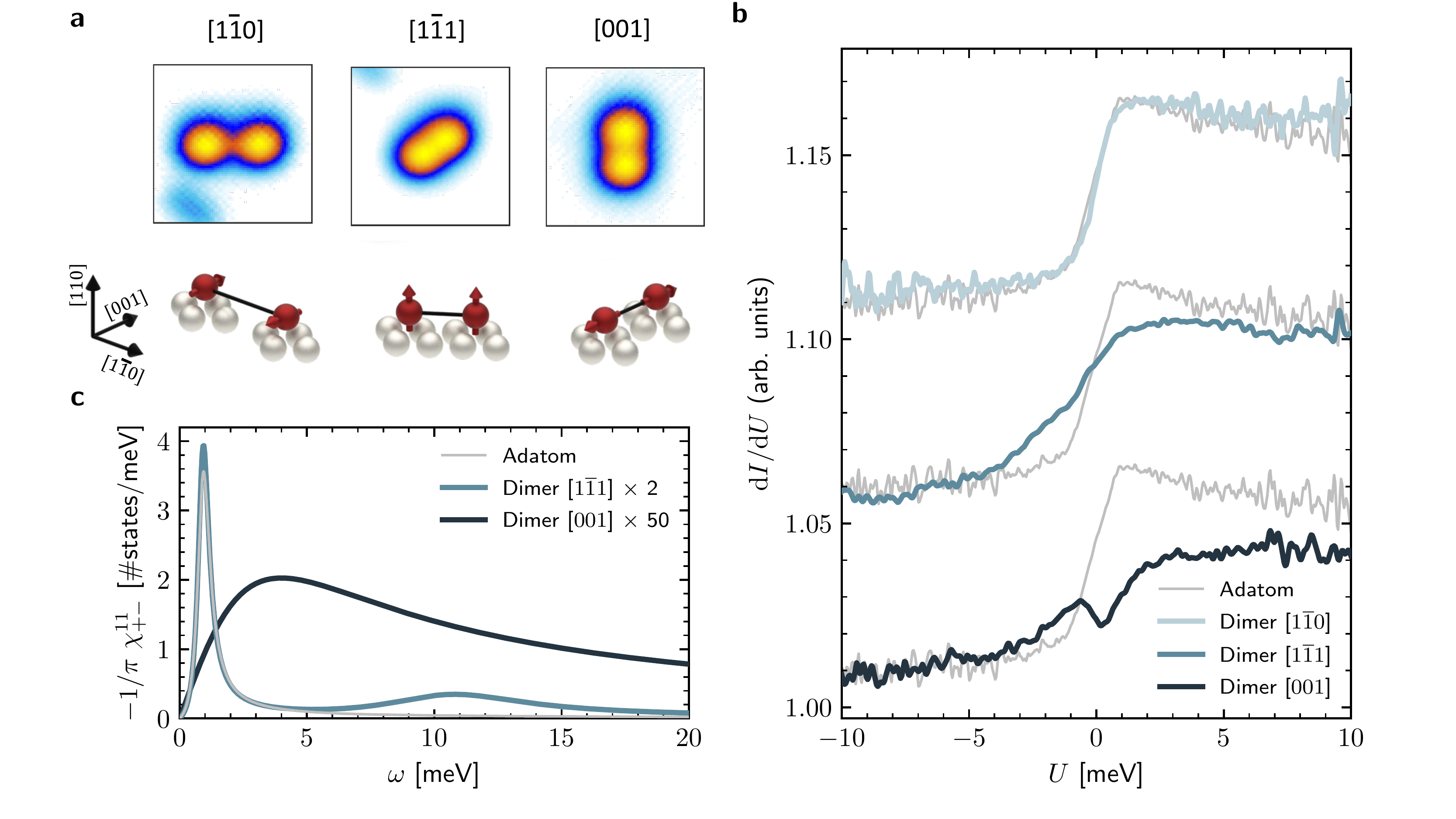}
        	\caption{
        	\textbf{Theoretical and experimental inelastic STS fingerprint of the Cr dimers along the [001], [1$\mathbf{\overline{1}}$0], and [1$\mathbf{\overline{1}}$1] directions on the Nb(110) surface.}
        	\textbf{a} Illustration of the dimers (STM topography)  and its grounds state magnetic structures.
        	The magnetic structures are obtained from self-consistent non-collinear DFT calculations. 
        	\textbf{b} Experimental inelastic STS spectra of the dimers, $I=\SI{3}{\nano\ampere}$ and $T=\SI{600}{\milli\kelvin}$ (coloured lines). 
        	The measurements on the single Cr adatom are shown as reference (thin grey lines, see also Fig.~\ref{fig:adatoms}).
        	The different spectra are vertically shifted to improve visibility. 
        	\textbf{c} Magnetic susceptibility obtained from TD-DFT. 
        	The spectra of the $[001]$ and $[1\overline{1}1]$ dimer are scaled by a factor of $50$ and $2$, respectively.
        	} \label{fig:dimers}
        \end{figure}
        \begin{table}
            \centering
            \begin{tabular}{l|ccccccc}
                 & Cr$_1$ & Cr$_2 [1\overline{1}0]$ & Cr$_2 [001]$ & Cr$_2 [1\overline{1}1]$ & Cr$_3 [1\overline{1}0]$ & Cr$_3 [001]$ & Cr$_3 [1\overline{1}1]$  \\ \hline
                 $M$ [\SI{}{\mub}] & {3.53} & {3.52} & {3.51} & {3.53} & {3.52} & {3.51} & {3.54} \\
                 $J_1$ [\SI{}{\milli\electronvolt}] & -- & {1.24} & {7.94} & {-8.95} & {1.31} & {7.97} & {-7.35} \\
                 $J_2$ [\SI{}{\milli\electronvolt}] & -- & -- & -- & -- & {-0.20} & {-0.31} & {-2.38} \\
                 $\MC{K}$ [\SI{}{\milli\electronvolt}] & {0.20} & {0.20} & {0.20}& {0.20} & {0.21} & {0.17} & {0.20}\\
            \end{tabular}
            \caption{
            \textbf{Ground state properties of the considered Cr-based nanostructures deposited on the Nb(110) surface.}
            Shown are the magnetic moments per atom $M$, which are averaged in case of the trimers, the next-nearest neighbor (NNN) isotropic exchange interaction $J_1$, the next NNN  isotropic exchange interaction $J_2$ along a given direction  
             and the magnetic anisotropy energy $K$, per atom. 
            Positive (negative) $J$ corresponds to an antiferromagnetic (FM) coupling, while a positive magnetic anisotropy energy indicates a preference for an in-plane orientation of the spin moments.
            }
            \label{tab:gs_prop}
        \end{table}
        To further explore the consistency of the experimental data with  the spin-excitation paradigm, we use atomic manipulation techniques to artificially create magnetic dimers oriented along different crystallographic directions on the Nb(110) surface. 
        We consider three next-nearest neighbor configurations: $[1\overline{1}0]$, $[1\overline{1}1]$, and $[001]$, illustrated in Fig.~\ref{fig:dimers}a. The highly anisotropic (110) surface  enables us  to create and scrutinize distinct direction-dependent magnetic ground states utilizing the coupling between the adatoms mediated by the conduction electrons.
        This effect is expected to be directly imprinted in the spin-excitation spectra which should show a clear direction-dependence. STS measurements confirm the existence of a strong direction-dependence behavior, as clearly visible in Fig.~\ref{fig:dimers}b. 
        A single step-like spectroscopic signature centered around the Fermi level is visible for all the three distinct directions. 
        However, a strong direction-dependent deviation with respect to the signal obtained for the isolated single Cr atom is evident, both in the intensity and in the broadening.  The $[1\overline{1}0]$ dimer is characterized by the largest separation and therefore the weakest coupling between the adatoms, resulting in only very minor differences with respect to the isolated atom case.
        In sharp contrast, both the $[1\overline{1}1]$ and the $[001]$ dimers show a significant broadening of the step-like feature. The spectrum corresponding to the $[001]$ dimer hosts a deep-like feature offset from zero-bias, surrounded by two maxima that are not symmetrically located with respect to the Fermi energy. In general, 
        these observations are in line with the theoretically calculated strength of the magnetic exchange interactions along different crystallographic directions summarized in Table \ref{tab:gs_prop}. 
        
        In the following, we present a theoretical analysis which allows us to effectively understand the origin of the distinct behaviour observed for dimers aligned along $[1\overline{1}1]$ and $[001]$ direction, where the magnetic coupling is the  strongest.  Since the adatoms are not nearest neighbors, the single ion anisotropy of each of the adatoms remains constant (see Table \ref{tab:gs_prop}). 
        However, the nature of the coupling is opposite along the two different directions, resulting in a strong antiferromagnetic (\SI{7.94}{\milli\electronvolt}) and ferromagnetic interaction (\SI{-8.95}{\milli\electronvolt}) for $[001]$ along the $[1\overline{1}1]$ directions, respectively. These predictions are in line with the experimental observations reported in Supplementary Note 11, where the magnetic coupling between next-nearest neighbours adatoms has been scrutinized by using a functionalized superconducting tip (a similar approach was introduced in Ref.~\onlinecite{Schneider2021}).  
        As demonstrated in the adatoms subsection, the analysis of the spin-excitations spectra can be carried out on the basis of the density of spin-flip excitations, which provide the intrinsic inelastic spectra. 
        These are obtained from the first-principles ground states illustrated in Fig.~\ref{fig:dimers}a resulting from the interplay of the single ion anisotropy, magnetic exchange interaction and magnetic field. 
        The density of spin-excitations are shown and compared to that of the isolated Cr adatom in Fig.~\ref{fig:dimers}c for each of the atoms forming the dimers.  
        
        Since the $[1\overline{1}1]$ dimer is ferromagnetic 
        (see Fig.~\ref{fig:dimers}a and Supplementary Note 7), the density of spin-excitation hosts a double peak structure, leading to a redistribution of the spectral weight observed for the isolated adatom. 
        They represent the acoustical and optical excitation modes, where in a classical picture the spin moments precess either clockwise or counter-clockwise, respectively. A simple description on the basis of a quantum spin model is given in Supplementary Note 9. 
        The position of the lowest mode, also observed experimentally, is dictated by the magnitude of the magnetic anisotropy energy ($\gamma' \MC{K} /M$) and shows a weaker intensity than that of the excitation mode of the isolated adatom with a broadening given by $\gamma' \alpha \MC{K}/M$ (See Supplementary Note 8). 
        The optical mode however occurs at energies  not resolvable within the experiment since it is dictated by the magnitude of the exchange interaction $\gamma' (\MC{K} - J) /M$ leading to a much broader feature ($\gamma' \alpha (\MC{K} -  J)/M$) than the acoustical one.

        In contrast to the ferromagnetic dimer, the antiferromagnetic one exhibits a single broad mode settled by the magnetic exchange interaction  ($\gamma' \sqrt{4\MC{K}  (\MC{K} +J) - \alpha^2 \left( J - 2 \MC{K} \right)^2}/2M$) since the excitation process promotes the dimer from a state where the total spin is 0 to 1. The broadening of the state is given by $\gamma' \alpha ( 2 \MC{K}  + J)/2M$, which explains the significantly large damping observed in both the experimental and theoretical spectra shown for the $[001]$ dimers in Fig.~\ref{fig:dimers}b-c.

        Overall, our theoretical analysis highlights that antiferromagnetic dimers are subjected to stronger electron-hole excitations, resulting in spin-excitations characterized by significantly larger energy broadening, i.e. shorter lifetimes, compared to the ferromagnetic case. This behavior manifests itself in a larger energy broadening along the $[001]$ direction as compared to the $[1\overline{1}1]$ direction, in agreement with our experimental observations. 
        
        Similarly to the adatom, we expect the spectra pertaining to the spin-excitation to result from a complex combination of excitation features hosted by majority and minority spin electrons at either sides of the bias voltage. This might explain the peculiar spectrum detected experimentally on the [001] dimer (see Fig.~\ref{fig:dimers}b), which hosts a  deep-like feature offset from zero-bias that is induced by the coupling of the two adatoms (see Supplementary Note 10). A similar kink is detected on the theoretical data of the adatom after application of a magnetic field (Figs.~\ref{fig:adatoms}e-f).

    \subsection{From trimers to long chains.}
        \begin{figure}
        	\centering
        	\includegraphics[width=\textwidth]{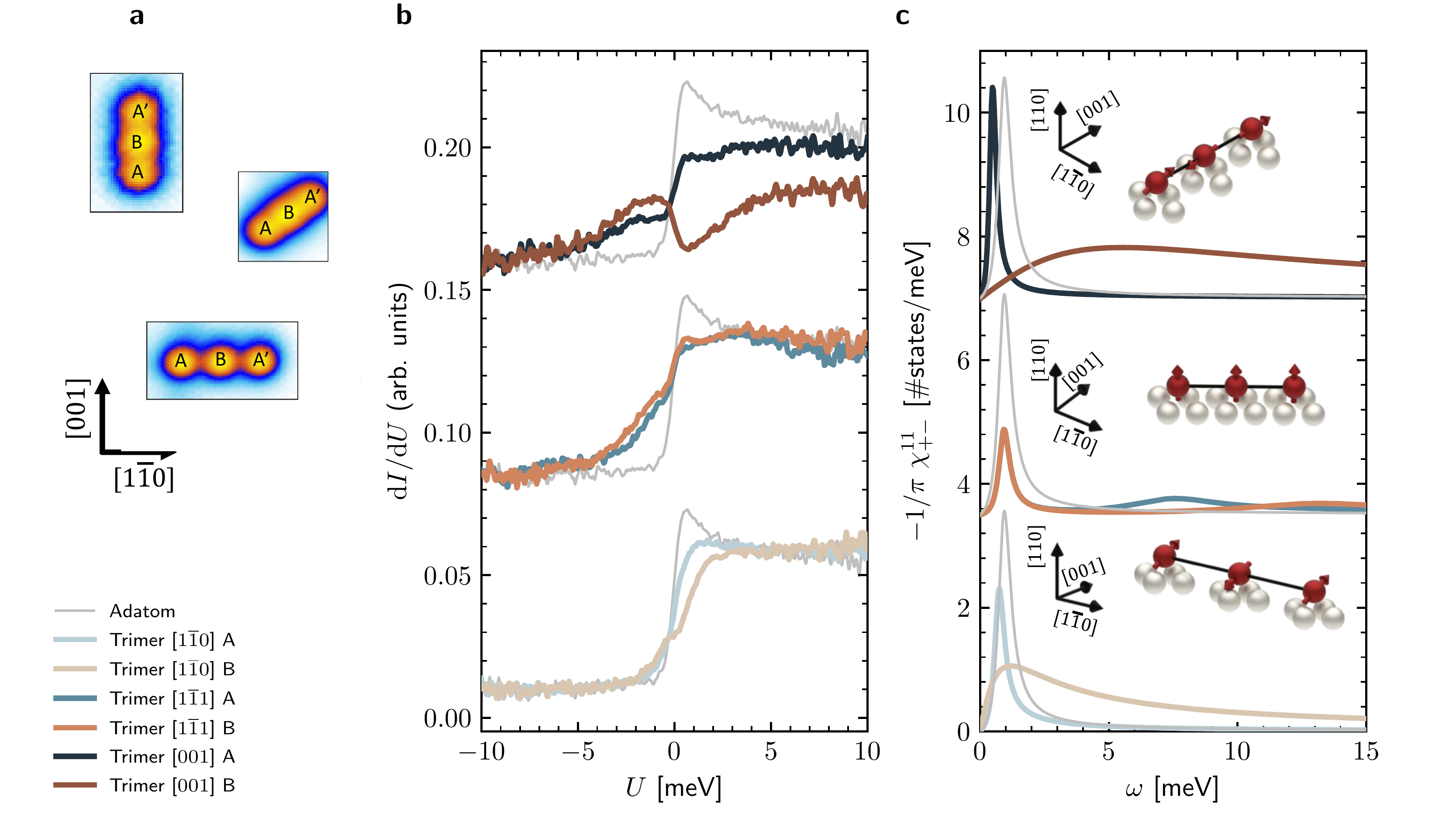}
        	\caption{
        	\textbf{Cr trimers along the [$\mathbf{001}$], [$\mathbf{1\overline{1}0}$], and [$\mathbf{1\overline{1}1}$] directions on the Nb(110) surface.}
        	\textbf{a} Illustration of the trimers (STM topography) and its grounds state magnetic structures at \SI{5}{\tesla}.
        	\textbf{b} Experimental inelastic STS spectra of the trimers at \SI{1}{\tesla} (coloured lines).
        	The measurements on the single Cr adatom are shown as reference (thin grey lines, see also Fig.~\ref{fig:adatoms}).
        	\textbf{c} Magnetic susceptibility obtained from TD-DFT at \SI{5}{\tesla}.
        	The spectra of atom B in the $[1\overline{1}0]$ direction and in the $[001]$ direction are scaled by a factor $10$ and $30$, respectively.
        	All trimer spectra are vertically shifted to improve visibility.
        	The insets show the magnetic structures obtained from self-consistent non-collinear DFT calculations.
        	} \label{fig:trimers}
        \end{figure}
        As identified for the dimers, the distinct spin-excitations spectra of the AFM and FM dimer is triggered by the alignment of their spin-moments. 
        To confirm this aspect, we artificially create a trimer along all distinct directions. 
        This breaks the perfect antiferromagnetic balance while it should not significantly affect the excitation behavior of the FM configurations. 
        STM images for all configurations are shown in Fig.~\ref{fig:trimers}a.
        Due to the $C_2$ symmetry, the two end atoms (A and A') can be distinguished from the central one (B) for all trimers.
        The inelastic STS spectra obtained at \SI{1}{\tesla} and \SI{600}{\milli\kelvin} for each Cr atom are shown in Fig.~\ref{fig:trimers}b.
        Similar to the dimers, a clear correlation between the strength of the magnetic coupling (see Table ~\ref{tab:gs_prop}) and the deviation of the inelastic STS signal from the single adatom case is visible. 
        Moreover, we also recover similar trends: the $[1\overline{1}0]$ trimer exhibits the weakest magnetic coupling and therefore only a weak deviation from the single adatom spectra is observed, while the observed broadening increases for the stronger coupled trimers along the $[1\overline{1}1]$ and $[001]$ directions.
        Noteworthy, we find two distinct trends depending on the nature of the magnetic coupling: The ferromagnetic trimer along the $[1\overline{1}1]$ direction shows only a very weak difference between the end atoms (A) and the central atom (B)  in contrast to the antiferromagnetic trimers, with the $[001]$ case hosting the largest discrepancy.
        
        To explain the observed spectra theoretically, we exploit the density of spin excitations of the trimers shown in Fig.~\ref{fig:trimers}c.
        For all trimers, the end atoms host a peak close to the excitation energy of the isolated adatom.
        The ferromagnetic $[1\overline{1}0]$ trimer shows peaks at higher energies depending on the nature of the considered atom. 
        Due to the increased coordination number of the central atom B, its high-energy peak is higher in energy than the one of atom A.
        For the antiferromagnetically coupled trimers along the $[001]$ and $[1\overline{1}0]$ directions, two fundamentally different spectra are observed depending on whether atom A or B is excited. 
        The low energy excitations contrast those observed for the dimers, which showed only a single broad feature. 
        The fundamental difference between the antiferromagnetic dimers and trimers is that the net magnetic moment cancels out in the former while it is finite for the latter. 
        This triggers a mixture of modes similar to those of the ferromagnetic and antiferromagnetic dimers. 
        Thus, unlike the dimers, the trimers can respond to an external magnetic field while preserving their collinear antiferromagnetic configuration. The lowest mode  corresponds to a collinear in-phase motion of all three atoms at an energy $\omega \approx 3 \gamma \MC{K} / M$ (see Supplementary Note 8), and the density of spin excitations in Fig.~\ref{fig:trimers}c shows that this mode is possible to be excited via the end atoms A and A'.
        In contrast, the central atom B only excites the modes at higher energy and with significantly increased broadening, i.e. translating to a larger coupling to electron-hole excitations, which explains to the observed inelastic STS signatures in Fig.~\ref{fig:trimers}b. By progressively increasing the size of the wires, we recover a spectroscopic behavior similar to that of the trimers (see Supplementary Note 12). 
        We note that a similar sequence of modes  was observed in Ref.~\onlinecite{Spinelli:2014}, which can equivalently be explained in terms of the nodal structure of the collective spin-excitations hosted by the nanomagnet.

\subsection{Spin-resolved spectroscopy.}
   \begin{figure}
    	\centering
     	\includegraphics[width=\textwidth]{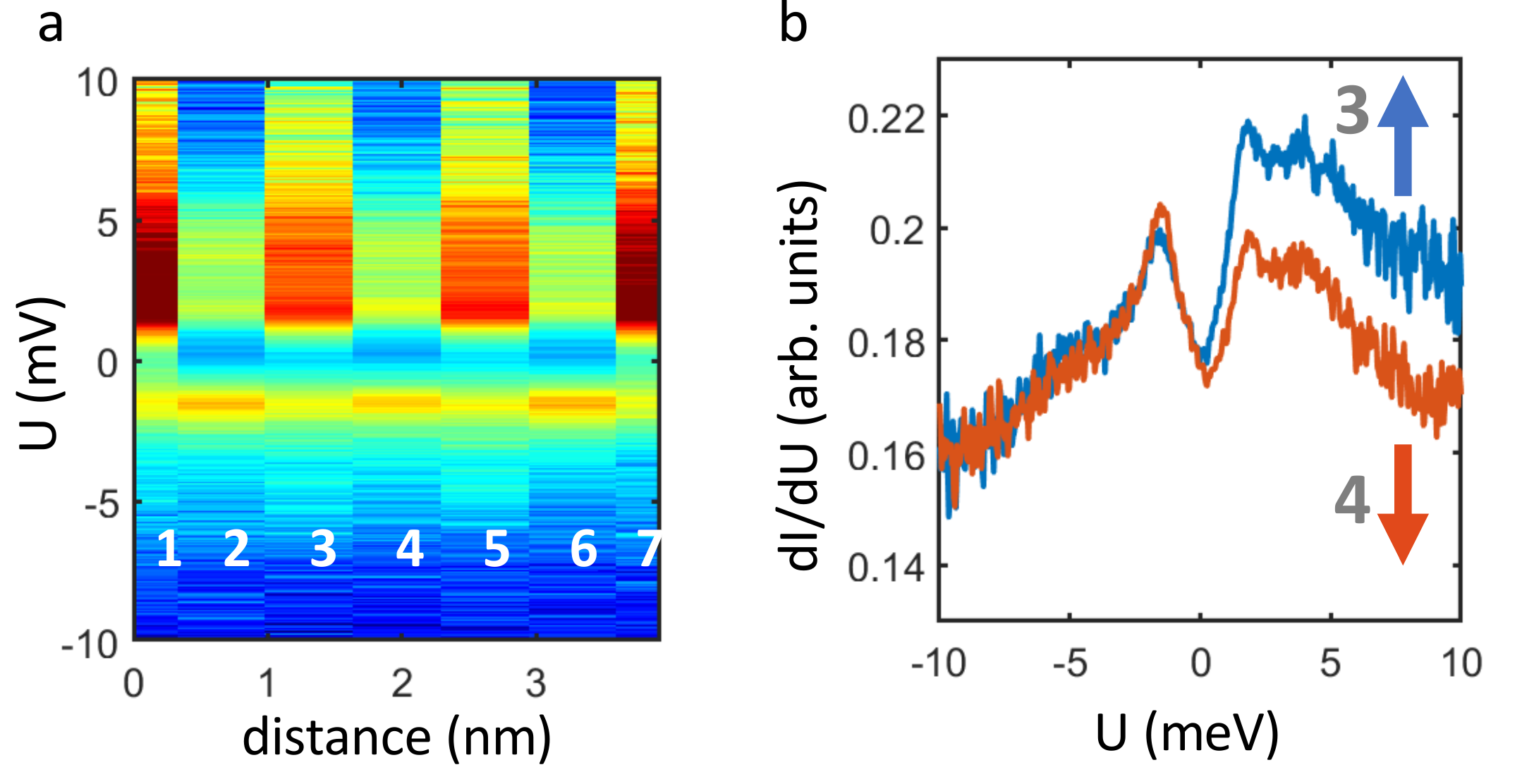}
      	\caption{
      	\textbf{Spin-polarized STS acquired on a 7 adatoms chain along  [001]. }
     	\textbf{a} Spectroscopic data collected by positioning the tip on top of each adatom along the chain (see numbers). The data reveal the existence of antiferromagnetic coupling. 
      	\textbf{b} Comparison between the STS spectra acquired on neighboring adatom (3 and 4 in \textbf{a}), highlighting the opposite spin contrast. Measurements have been acquired at \SI{5}{\tesla} with a Cr-terminated tip. Stabilization parameters: $V$ = -10 meV, $I$ = 3 nA, and modulation $V_{rms}$ = 200 $\mu$V. 
        	} \label{fig:spin}
        \end{figure}
Finally, we experimentally scrutinize the spin-resolved spectroscopic signatures for chains oriented along the [001], i.e. when neighboring adatoms are strongly antiferromagnetically coupled (see also related Supplementary Notes 11 and 12). This is a particularly interesting scenario, since the splitting of the zero bias anomaly  is the strongest (see section on dimers). This is expected to result in well-separated spectroscopic signatures for opposite spin channels, a direct result of the magnetic interaction between the adatoms which renormalizes the spectrum while simultaneously removing the spin degeneracy \cite{PhysRevLett.108.087203}. This situation is analogue to the theoretical plots presented in Fig. 1f for a single adatom. Indeed, the coupling between adatoms can be considered as an effective magnetic field. Figure 4 a shows the spin resolved spectroscopic data acquired with a Cr-terminated W tip under a magnetic field strength of \SI{5}{\tesla} applied perpendicular to the sample surface. Spectra obtained by positioning the tip on top of each one of the adatom in the chain are indicated by numbers. An alternating contrast is clearly visible along the chain, with a slightly different signature for adatoms localized at the end of the chain (number 1 and 7 in panel a), a direct consequence of their different local environment compared to adatoms in the bulk of the chain. The spectral shape is better highlighted in Fig. 4b, which directly compares the spectra acquired on antiferromagnetically coupled adatoms i.e. number 3 and 4 in Fig. 4a. The spectra for opposite spin channels are in very good agreement with the theoretically calculated spectra shown in Fig. 1f. i.e. a relatively sharp peak center below the Fermi level for the majority-spin channel, and a broader step-like signature at positive energy for minority spin electrons.
        
\section*{Discussion}
    In conclusion, we systematically unveiled zero-bias anomalies in Cr atom nanostructures deposited on the metallic Nb(110) surface. Our inelastic STS measurments reveal an asymmetric single step-like zero bias feature, which is preserved but can be modulated depending on the strength of the external magnetic field, the size of the nanostructure, and its spatial orientation on the substrate. The latter is  used to engineer inter-atom magnetic coupling to be ferromagnetic or antiferromagnetic. We demonstrated the possibility of tracking these complex spectral line shapes through ab-initio simulations combining time-dependent density functional and many-body perturbation theories, evidencing that by constructing the nano-object with the desired magnetic state, one can control the underlying excitation energies and related lifetimes by impacting on
the number of emitted electron-hole excitations. The emitted spin-currents accompanying the electron-hole excitations characterizing each excitation mode might prove useful for spin-pumping nanoscale devices to explore innovative spintronics concepts based on spin-transfer/spin-orbit torques and spin-Hall effects, for the manipulation of quantum spins.
Our results shed new light on the interpretation of zero-bias anomalies and pave a way to engineering low-energy features by manipulating the spin excitation spectra by utilizing the underlying magnetic structure.

\begin{methods}

\subsection{Sample and tip preparation.}
            
    To obtain an atomically flat and clean substrate, a commercial Nb single crystal polished on one side with (110) surface normal was heated up to \SI{2300}{\kelvin} multiple times by short flashes of electron bombardment and under UHV conditions~\cite{Odobesko2019}. The procedure is repeated until the main contaminant oxygen gives way to large areas of clean Nb. In-situ deposition of Cr atoms is done by keeping the substrate cold below \SI{15}{\kelvin} to prevent migration and to avoid clustering of adsorbed single atoms. A commercial STM head from Scienta Omicron is employed to carry out topographic imaging, scanning tunneling spectroscopy measurements and atomic manipulation of Cr adatoms. Data is acquired in $10^{-10}\si{\milli\bar}$ pressure at a temperare of \SI{600}{\milli\kelvin} and with an applied external magnetic field of $>$\SI{1}{\tesla} using an etched tungsten tip characterized on Ag(111). Bringing the tip close to the adatoms by increasing the constant tunneling current set point to about \SI{70}{\nano\ampere} at a bias voltage of \SI{-5}{\milli\volt} enabled us to drag the Cr single atoms to the desired lattice positions and to isolate all the nanostructures under investigation from surrounding Cr atoms by at least \SI{4}{\nano\meter}.

\subsection{First-principles calculations.}

    The methodology followed in our work is based on a first-principles approach as implemented in the framework of the scalar-relativistic full-electron Korringa-Kohn-Rostoker (KKR) Green function augmented self-consistently with spin-orbit interaction~\cite{Papanikolaou:2002,Bauer:2014}. The spin-excitations are described in a formalism based on time-dependent density functional theory (TD-DFT)~\cite{Lounis:2010,Lounis:2011,Lounis:2015,Manuel:2015} including spin-orbit interaction. 
    Many-body effects triggered by the presence of spin-excitations are approached via many-body perturbation theory~\cite{Schweflinghaus:2014,Schweflinghaus:2016,Ibanez:2017} extended to account for relativistic effects \cite{Bouaziz:2020a}. The utilized  Multiple-scattering theory enables an embedding scheme, which is versatile for the treatment of nanostructures in real space. The local spin density approximation (LSDA) is employed for the evaluation of the exchange-correlation potential~\cite{Vosko:1980}
    while the full charge density is computed within the atomic-sphere approximation (ASA).  
    We assume an angular momentum cutoff at $\ell_{\text{max}} = 3$ for the orbital expansion of the Green function and when extracting the local density of states a k-mesh of $450 \times 450$ is considered.   
    The Cr atoms sit on the hollow stacking site relaxed towards the surface by $20\%$ of the interlayer distance of the underlying Nb(110) surface, as found in previous studies~\cite{Kuester2021}.
    
    The single-particle Green functions corresponding to the ground-state are  employed for the construction of the tensor of dynamical magnetic susceptibilities, $\underline{\chi}(\omega)$, within  time-dependent density functional theory (TD-DFT)~\cite{Lounis:2010,Lounis:2015,Manuel:2015} including spin-orbit interaction. In practice, we solve the Dyson-like equation relating the full susceptibility to the Kohn-Sham susceptibility, $\underline{\chi}_\text{KS}(\omega)$ as
    \begin{equation}
    \underline{\chi}(\omega) = \underline{\chi}_\text{KS}(\omega) 
    + \underline{\chi}_\text{KS}(\omega)\,\underline{\mathcal{K}}\,\underline{\chi}(\omega) \quad ,
    \label{TDDFT_RPA}
    \end{equation}
    where $\underline{\chi}_\text{KS}(\omega)$ describes uncorrelated electron-hole excitations, while $\underline{\mathcal{K}}$ represents the exchange-correlation kernel, taken in adiabatic LSDA (such that this quantity is local in space and frequency-independent~\cite{Gross:1985}). 
    The energy gap in the spin excitation spectrum is accurately evaluated using a magnetization sum rule~\cite{Lounis:2010,Lounis:2015,Manuel:2015}. 
    \\
    The magnetic exchange interactions in the dimers and trimers were obtained using the magnetic force theorem in
    the frozen-potential approximation and the infinitesimal rotation method \cite{Ebert2009}.
    The on-site magnetic anisotropy of the Cr atoms in all systems is  obtained from the method of constraining fields \cite{Brinker2019}. 
    More details are given in Supplementary Notes 6 and 7.
    
    Considering the reasonable theoretical recovery of the experimentally observed low-energy anomaly, we discuss here  the impact of the assumptions made within the simulations. It is clear that the assumed LSDA is performing well in describing the magnitude of the magnetic anisotropy energy, which dictates the position of the spin-excitation, and of the magnetic exchange interactions since the spin-polarized STS experiments are in nice agreement with the ab-initio simulations. However, it is very possible that a slight shift in the magnetic anisotropy energy and of the electronic states at the Fermi energy, which defines the lifetime/broadening of the zero-bias anomalies, by changing the exchange and correlation potential could improve the agreement between theory and experiment.  
\end{methods}

\begin{addendum}
\item  S.B. and S.L.~thank M. dos Santos Dias,  A. M. Montero, Filipe S. M. Guimar\~aes for fruitful discussions. \textbf{Funding:} This work is supported by the European Research Council (ERC) under the European Union's Horizon 2020 research and innovation programme (ERC-consolidator grant 681405 — DYNASORE). We acknowledge the computing time granted by the JARA-HPC Vergabegremium and VSR commission on the supercomputer JURECA at Forschungszentrum Jülich. 
 \textbf{Authors contributions:} P.S. and S.L. initiated, designed and supervised the project. F.K. performed all the STM measurements supervised by S.S.P and P.S.. S.B. performed the ab-initio simulations supervised by S.L.. S.B. and S.L. wrote the manuscript to which all authors contributed via discussions and corrections. \textbf{Competing Interests:} The authors declare that they have no competing interests. \textbf{Data and materials availability:} All data needed to evaluate the conclusions in the paper are present in the paper. 
\item[Supplementary Materials.] This PDF file includes:

Sections S1 to S12\\
Figs. S1 to S10\\
Tables S1 to S2 

\end{addendum}

\section*{References and notes}

\bibliographystyle{Science}
\bibliography{mylib_paper.bib}



\section*{Supplementary material}

# Supplementary Materials for

## Anomalous excitations of atomically crafted quantum magnets

Sascha Brinker*, Felix Küster, Stuart S. P. Parkin, Paolo Sessi, Samir Lounis*

*Corresponding author. Email: s.brinker@fz-juelich.de, s.lounis@fz-juelich.de

**This PDF file includes:**

Sections S1 to S12
Figs. S1 to S10
Tables S1 to S2

1

# Supplementary Note 1:   Electronic structure – local density of states

The local density of states of the considered systems as is shown in Supplementary Figure S1.

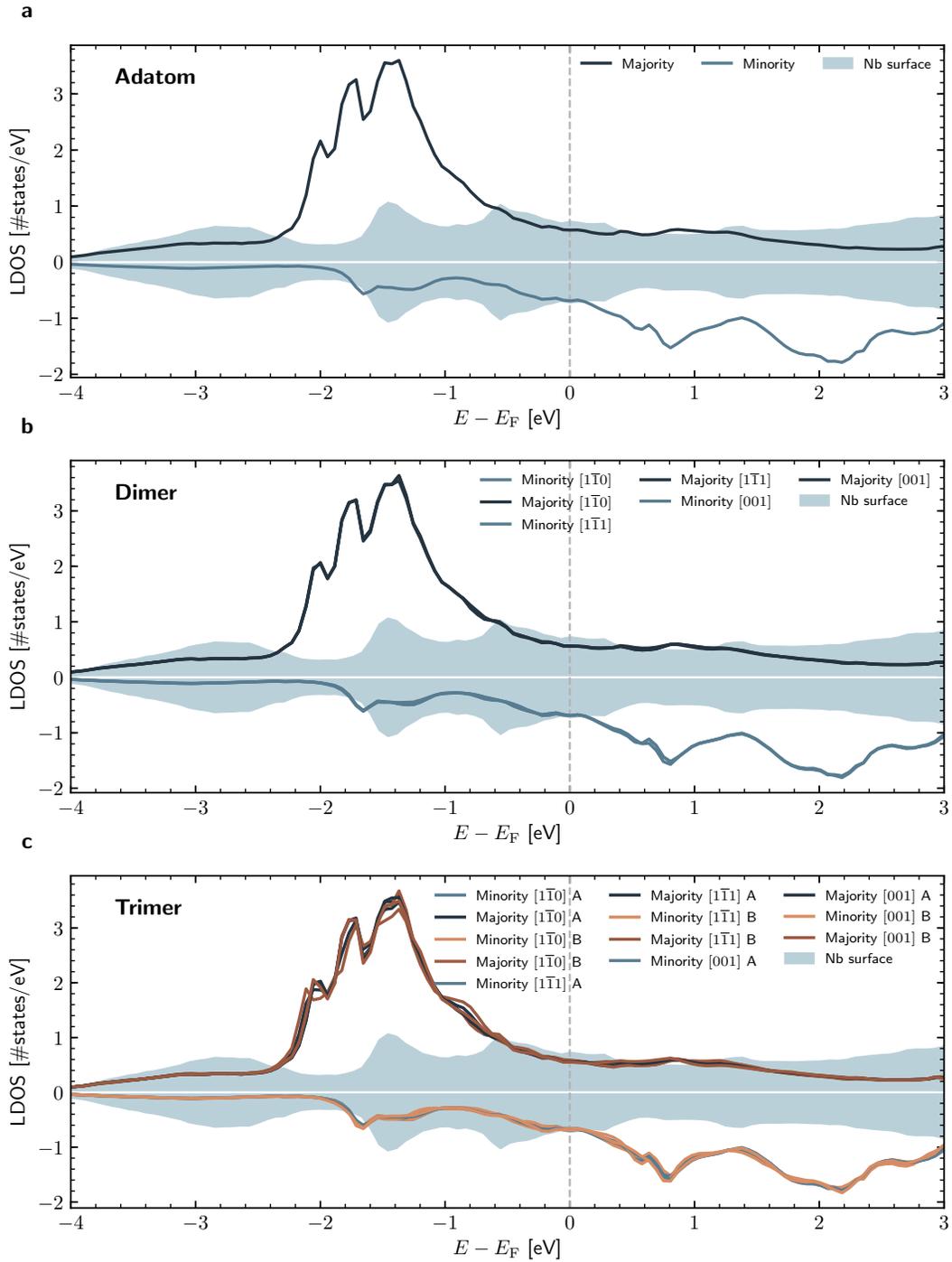

**Figure S1: Local density of states (LDOS) of the different considered systems.** The filled background shows the LDOS of the first Nb surface layer, while the lines show the LDOS of the adatom (**a**), the dimers (**b**), and the trimers (**c**).

# Supplementary Note 2: Fano shape fit of the zero-bias anomaly

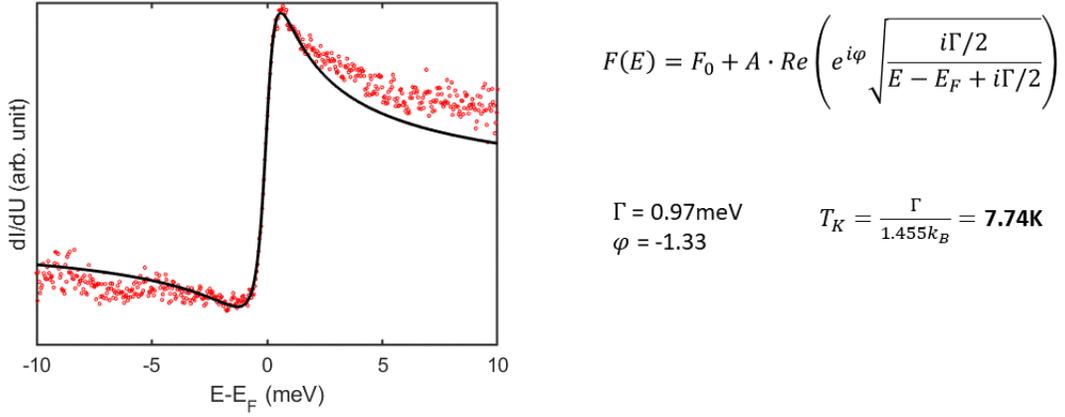

$$F(E) = F_0 + A \cdot Re\left(e^{i\varphi}\sqrt{\frac{i\Gamma/2}{E - E_F + i\Gamma/2}}\right)$$

$\Gamma$ = 0.97meV
$\varphi$ = -1.33

$T_K = \dfrac{\Gamma}{1.455 k_B} =$ **7.74K**

**Figure S2: Fit of the zero-bias anomaly for a single Cr adatom according to a Frota function.** The red dots correspond to the experimental spectrum while the black line corresponds to the fit. To minimize the influence of the tip density of states, the spectrum has been normalized to the one acquired by positioning the tip over the bare Nb(110) substrate using the very same W microtip. Measurements have been acquired under a magnetic field of $B = 2$ T applied perpendicular to the sample surface to drive the Nb substrate into the metallic regime.

# Supplementary Note 3: Measurements at different set-points currents

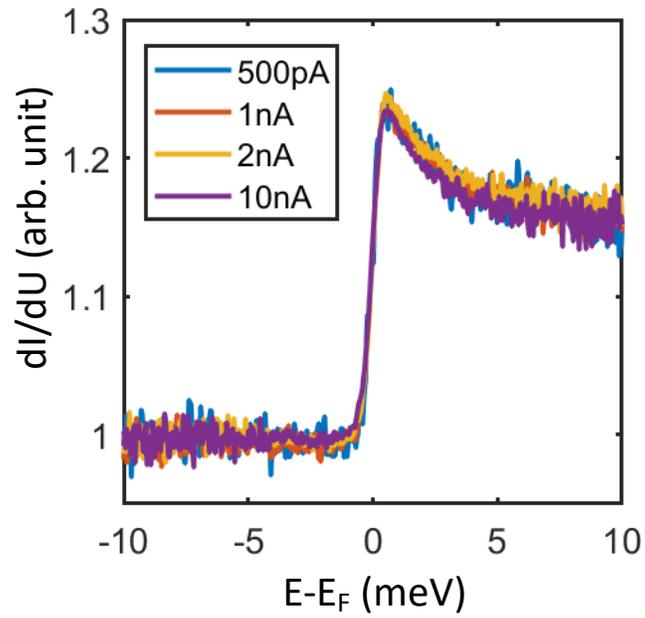

**Figure S3: Zero-bias anomaly for a single Cr adatom acquired by stabilizing the tip at progressively higher tunneling currents.** The zero bias anomaly does not show any spectral change by increasing the current.

# Supplementary Note 4: Spatial mapping of the zero-bias anomaly

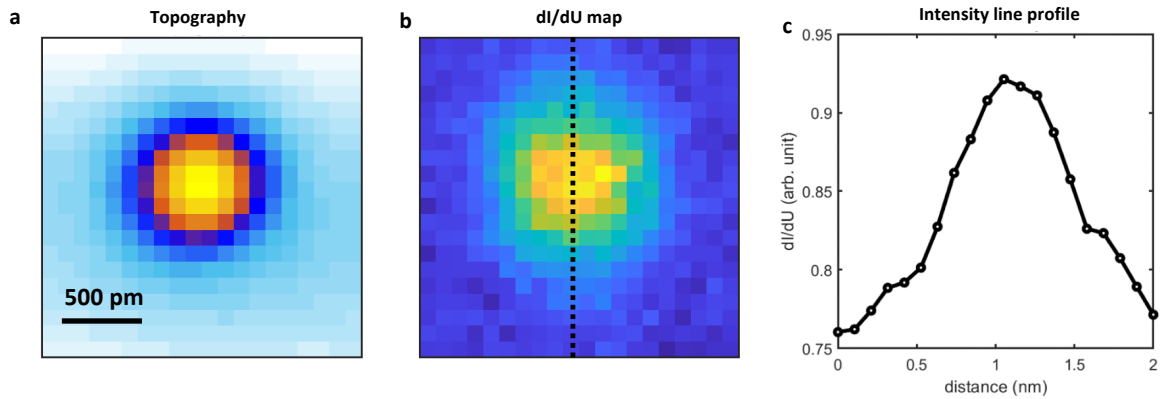

Figure S4: **Spatial mapping of the zero-bias anomaly of Cr adatom deposited on the Nb(110) surface.** **a** Topography of the probed single Cr adatom. **b** Spatial mapping of the zero-bias anomaly acquired by a full spectroscopic grid. The map corresponds to the intensity of the step-like spectroscopic signature at the energy E = 0.750 $\mu$V. **c** Corresponding line profile of the intensity along the black dashed line in **b**. The data allow to visualize that the zero-bias anomaly is strongly localized onto the Cr adatom.

# Supplementary Note 5: Zero bias anomaly on Mn

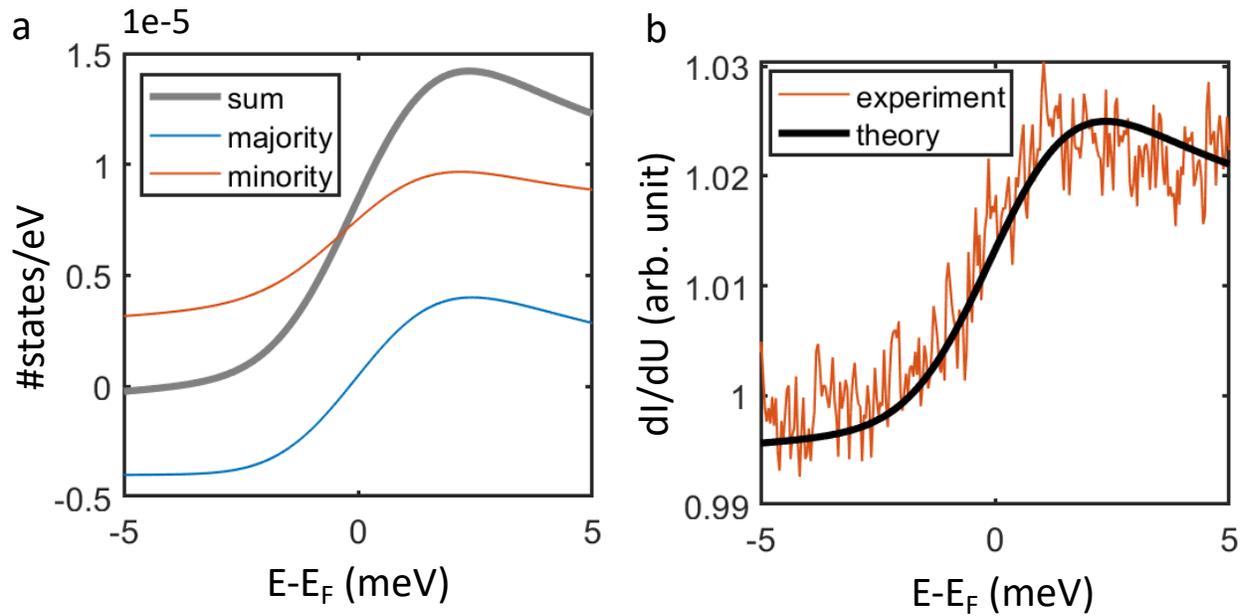

**Figure S5: Zero-bias anomaly for a single Mn adatom coupled to the Nb(110) surface.** Similarly to Cr, a step-like spectroscopic signature centered at the Fermi level is found also for Mn as obtained theoretically from ab-initio **a** and experimentally **b**. Compared to Cr, the Mn signature is broader and weaker in intensity. As evident by comparing panels **a** and **b**, also for Mn the theoretically calculated spectra can describe the experimental results with high accuracy.

# Supplementary Note 6: Symmetry analysis of the magnetic interactions

To analyse the ab-initio results, we use throughout this paper a generalized Heisenberg model,

$$\mathcal{H} = \sum_i \mathbf{e}_i \mathcal{K}_i \mathbf{e}_i + \frac{1}{2} \sum_{ij} J_{ij} \mathbf{e}_i \cdot \mathbf{e}_j + \frac{1}{2} \sum_{ij} \mathbf{D}_{ij} \cdot \mathbf{e}_i \times \mathbf{e}_j + \frac{1}{2} \sum_{ij} \mathbf{e}_i J_{ij}^{\text{sym}} \mathbf{e}_j - \sum_i m_i \mathbf{B}_i \cdot \mathbf{e}_i \quad , \tag{S.1}$$

where $i$ labels the Cr atoms with $\mathbf{e}_i$ being the direction of its magnetic moment, $\mathcal{K}_i$ is the magnetic on-site anisotropy, $J_{ij}$ is the isotropic exchange interaction, $\mathbf{D}_{ij}$ is the Dzyaloshinskii-Moriya interaction, $J_{ij}^{\text{sym}}$ is the symmetric anisotropic exchange interaction, and $\mathbf{B}_i$ is an external magnetic field. Depending on the symmetry of the system the allowed form of the interactions can vary.

**Adatom in $C_{2v}$ symmetry.** The isolated adatom follows the $C_{2v}$ symmetry of the underlying Nb(110) lattice, which results in a magnetic anisotropy matrix of the following form:

$$\mathcal{K} = \begin{pmatrix} \mathcal{K}_x & 0 & 0 \\ 0 & \mathcal{K}_y & 0 \\ 0 & 0 & 0 \end{pmatrix} \quad . \tag{S.2}$$

Note that the $zz$-component was set to 0, which results from the freedom of the trace of the magnetic anisotropy.

**Dimer in $C_{2v}$ and $C_2$ symmetry.** The Cr dimers oriented in the high-symmetry directions [001] and [1$\bar{1}$0] preserve the $C_{2v}$ symmetry of the underlying lattice, while the dimer in the [1$\bar{1}$1] has the reduced symmetry $C_2$. Even though the magnetic on-site anisotropy is in general complex with off-diagonal elements (due to the lack of local symmetries, at most one mirror plane is present for the discussed dimers), we assume it to be similar to the one of the isolated adatom since we discuss only next-nearest neighbor dimers resulting in a weak coupling and only a weak renormalization of the on-site anisotropy is expected. This assumption is supported by the band energy differences in the frozen potential approximation shown in Table 1 of the main text, which do not show any siginificant renormalizations of the anisotropy when dimers and trimers are built.
In the general case only a rotation along the $z$ axis by 180° is present leading to the symmetry constraints,

$$\mathbf{D}_{12} = (D_x, D_y, 0) \qquad \text{and} \qquad J_{12}^{\text{sym}} = \begin{pmatrix} J_{xx} & J_{xy} & 0 \\ J_{xy} & J_{yy} & 0 \\ 0 & 0 & -J_{xx} - J_{yy} \end{pmatrix} \quad , \tag{S.3}$$

while $J_{xy} = 0$, and $D_y = 0$ and $D_x = 0$, respectively, for the dimers in the high-symmetry directions [001] and [1$\bar{1}$0].

**Trimer in the $C_{2v}$ and $C_2$ symmetry.** Similarly to the dimers the trimers can be grouped into the trimers preserving the $C_{2v}$ symmetry of the lattice (those along high-symmetry directions) and the [1$\bar{1}$1] trimer which reduces the symmetry to $C_2$. The magnetic on-site anisotropy is assumed to be similar to the one of the isolated adatom. For the trimers there are two possible types of interactions connecting either direct neighbors (atoms 1 and 2 and atoms 2 and 3) or the end atoms of the trimer (atoms 1 and 3). The interactions of the former pairs can always be related to each other using the rotational symmetry along the z-axis present in $C_2$ and $C_{2v}$. These interactions simplify further for the trimers along the high-symmetry directions leading $D_y$ and $D_x$ as the only finite off-diagonal components for the [001] and the [1$\bar{1}$0] direction, respectively. The interaction between the two end atoms of each trimer follows the same symmetry rules as the one of the dimers in the respective directions.

## Supplementary Note 7: Magnetic ground states

The magnetic ground states of the systems can be either obtained by minimizing the Heisenberg model, eq. (S.1), with parameters obtained from DFT (see Methods of the main text, Table 1 of the main text, and Supplementary Table S1), or by calculating the non-collinear ground state self-consistently from DFT. The latter approach is computationally more expensive, but also more precise, since it includes for example potential higher-order contributions to the magnetic anisotropy or the magnetic exchange interactions (see e.g. Ref. [69]). We used this approach to obtain the magnetic ground states of all systems, which are shown in Supplementary Table S2.

**Adatom.** The magnetic moment in the zero field ground state of the adatom is parallel to the $y$-direction, which is favoured by the magnetic on-site anisotropy,

$$\mathcal{K} = \begin{pmatrix} -0.188 & 0.0 & 0.0 \\ 0.0 & -0.200 & 0.0 \\ 0.0 & 0.0 & 0.0 \end{pmatrix} \mathrm{meV} \quad . \tag{S.4}$$

When a field is applied in the out-of-plane direction ($z$-direction), our self-consistent first-principles calculations indicate that the adatom starts to tilt towards the out-of-plane direction, and for fields $\geq 3\,\mathrm{T}$ it fully aligns with the field.

**Dimer.** The ferromagnetic dimer in the $[1\bar{1}1]$ direction almost aligns with the magnetic field exhibiting a small tilt of 2.4° in the polar angles due to the DMI (see Supplementary Table S1). The other dimers stay mostly in their AFM in-plane ground state with only a weak impact of the field. Noteworthy, the [001] dimer even tilts towards the $-z$-direction counteracting the field, which is due to a competition with the relatively strong DMI in the system.

**Trimer.** The ferromagnetic $[1\bar{1}1]$ trimer aligns with the field with a small opening of 4.3°. Compared to the previously discussed antiferromagnetic dimers, the antiferromagnetic trimers show a quite different behaviour. Due to the odd number of atoms, the trimers exhibit a finite net magnetic moment, and can therefore respond to a magnetic field without competing with the internal strong AFM exchange. If we ignore the internal spin structrue, they act like a macrospin with the same magnetic moment as a single adatom but three times larger magnetic anistropy. Therefore, the weak AFM trimer in the $[1\bar{1}0]$ direction tilts towards the out-of-plane direction with an effective polar angle of 64.7°. The strong AFM trimer in the [001] direction shows additional cantings induced by a strong DMI.

|  | $\mathrm{Cr}_2[1\bar{1}0]$ | $\mathrm{Cr}_2[001]$ | $\mathrm{Cr}_2[1\bar{1}1]$ | $\mathrm{Cr}_3[1\bar{1}0]$ | $\mathrm{Cr}_3[001]$ | $\mathrm{Cr}_3[1\bar{1}1]$ |
|---|---|---|---|---|---|---|
| $|\mathbf{D}_1|$ [meV] | 0.05 meV | 0.21 meV | 0.84 meV | 0.05 meV | 0.33 meV | 0.81 meV |

**Table S1: Magnitude of the next nearest-neighbor Dzyaloshinskii-Moriya interaction.** The magnetic interaction $|\mathbf{D}_1|$ among the closest neighbors is stable with the size of the nanostructures (Cr atoms are not neareast neighbors) but changes dramatically with their crystallographic direction.

| System | Polar angles (scf) | Azimuthal angles (scf) |
|---|---|---|
| Adatom (0 T) | 90° | 90° |
| Adatom (1 T) | 64° | 90° |
| Adatom (2 T) | 30° | 90° |
| Adatom (≥3 T) | 0° | 0° |
| $[1\bar{1}1]$ dimer (5 T) | $(2.4°, 2.4°)$ | $(-66.5°, 113.5°)$ |
| $[1\bar{1}0]$ dimer (5 T) | $(87.5°, 87.5°)$ | $(90°, -90°)$ |
| $[001]$ dimer (5 T) | $(91.3°, 91.3°)$ | $(90°, -90°)$ |
| $[1\bar{1}1]$ trimer (5 T) | $(4.3°, 0.0°, 4.3°)$ | $(-72.5°, -101.9°, 107.5°)$ |
| $[1\bar{1}0]$ trimer (5 T) | $(64.7°, 113.5°, 64.7°)$ | $(89.6°, -90.0°, 90.4°)$ |
| $[001]$ trimer (5 T) | $(86.4°, 98.3°, 76.0°)$ | $(90.0°, -90.0°, 90.0°)$ |

**Table S2: Magnetic ground state of the adatom, dimers and trimers on the Nb(110) surface obtained self-consistently from DFT.** The polar and azimuthal angles define the direction of the magnetic moments carried by the adatoms.

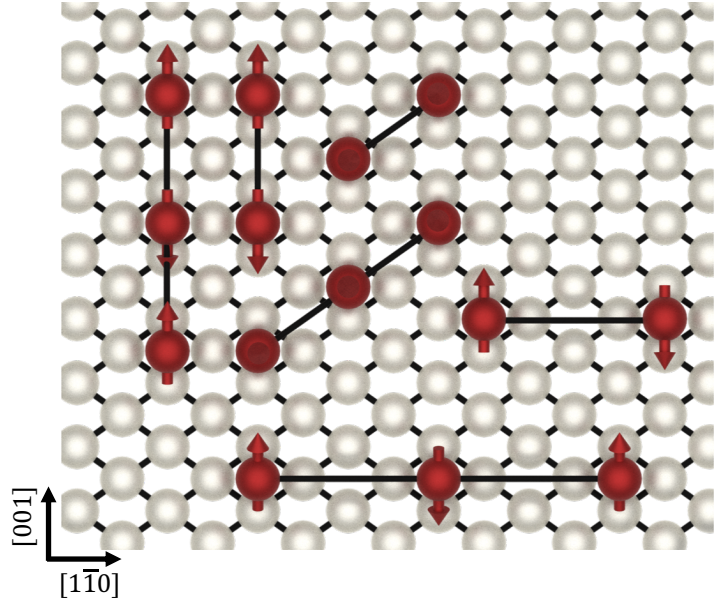

**Figure S6: Illustration of the magnetic ground states of all dimers and trimers on the Nb(110) surface**. An out-of-plane magnetic field of 5 T is assumed. The Cr nanostructures are represented by red spheres with arrows connected by black lines while the Nb atoms are shown as grey spheres. The corresponding polar and azimuthal angles are given in Supplementary Table S2.

## Supplementary Note 8: Landau-Lifshitz-Gilbert model

The Landau-Lifshitz-Gilbert (LLG) model [55,56] describes the transversal dynamics of magnetic moments, and can in general be written as

$$\frac{d\mathbf{m}_i}{dt} = -\gamma \mathbf{m}_i \times \left( \mathbf{B}_i^{\text{eff}} + \sum_j \mathcal{G}_{ij} \cdot \frac{d\mathbf{m}_j}{dt} \right) \quad , \tag{S.5}$$

where $\gamma$ is the gyromagnetic ratio, $\mathbf{B}_i^{\text{eff}}$ is the effective field acting on the moment $i$ and $\mathcal{G}_{ij}$ is the Gilbert damping tensor. The effective field can be obtained using any convenient spin model like the generalized Heisenberg model discussed in Supplementary Note 6:. It can be shown that in the LLG model the transversal magnetic susceptibility can be calculated by

$$\chi_{i\alpha j\beta}^{-1}(\omega) = \delta_{ij} \left( \delta_{\alpha\beta} \frac{B_{iz}^{\text{eff}}}{M_i} + \frac{i\omega}{\gamma M_i} \epsilon_{\alpha\beta\mu} \right) + \frac{1}{M_i M_j} (\mathcal{R}_i J_{ij} \mathcal{R}_j^\mathsf{T})_{\alpha\beta} + i\omega (\mathcal{R}_i \mathcal{G}_{ij} \mathcal{R}_j^\mathsf{T})_{\alpha\beta} \quad , \tag{S.6}$$

where $\mathcal{R}_i$ describes a rotation to the local spin frame of the magnetic moment $i$, which is oriented such that the local $z$-coordinate aligns with the direction of the magnetic moment. In the following, we will approximate the Gilbert damping tensor by $\mathcal{G}_{ij} = \delta_{ij}\alpha/\gamma m_i$, where $\alpha$ is the isotropic Gilbert damping parameter. The equally important off-diagonal elements of the Gilbert damping tensor can be shown to renormalize the gyromagnetic ratio and can thus be neglected if we assume a general gyromagnetic ratio with $\gamma \neq 2$.

**Adatoms in a magnetic field** The first example we discuss is the single adatom in an external magnetic field. We assume an out-of-plane orientation of the magnetic moment of the adatom, which holds for our system for fields $\gtrsim 2\,\mathrm{T}$ (see Supplementary Note 7:). Using the spin model of the single adatom discussed in eqs. (S.1) and (S.2) and the definition of the inverse transverse susceptiblity in the LLG model given in eq. (S.6) one finds,

$$\chi^{-1}(\omega) = \begin{pmatrix} \frac{K_x}{M^2} + \frac{B}{M} + i\frac{\alpha\omega}{\gamma M} & i\frac{\omega}{\gamma M} \\ -i\frac{\omega}{\gamma M} & \frac{K_y}{M^2} + \frac{B}{M} + i\frac{\alpha\omega}{\gamma M} \end{pmatrix} \quad , \tag{S.7}$$

with the eigenfrequencies given by $\det(\chi^{-1}(\omega)) = 0$ leading to,

$$\omega_\pm = \frac{\gamma}{(1+\alpha^2)} \frac{1}{2M} \left( i\alpha(K_x + K_y + 2BM) \pm \sqrt{4(K_x + BM)(K_y + BM) - \alpha^2 (K_x - K_y)^2} \right) \quad . \tag{S.8}$$

Defining $\gamma' = \frac{\gamma}{1+\alpha^2}$ the broadening (defining the FWHM of the excitations) of the eigenmodes is given by

$$\mathrm{Im}(\omega_\pm) = \alpha\gamma' \frac{K_x + K_y + 2BM}{2M} \quad , \tag{S.9}$$

while the energies of the eigenmodes are given by,

$$\mathrm{Re}(\omega_\pm) = \pm \frac{\gamma'}{2M} \sqrt{4(K_x + BM)(K_y + BM) - \alpha^2(K_x - K_y)^2} \quad . \tag{S.10}$$

Note that in the main manuscript we simplify the equations by using $K_x = K_y$, which is a very good approximation for the discussed Cr adatom (see eq. S.4), and consequently also for the dimers and trimers.

**Collinear dimers** Considering only on-site anisotropies and isotropic exchange interactions, the effective fields are given by

$$\mathbf{B}_i^{\text{FM/AFM}} = \mp \frac{\mathbf{e}_z}{M} J \quad , \quad i = \{1, 2\} \quad . \tag{S.11}$$

Assuming a local and isotropic Gilbert damping tensor, $\mathcal{G}_{ij,\alpha\beta} = \delta_{ij}\delta_{\alpha\beta}\alpha$, one finds for the susceptibility of the ferromagnetic dimer in the $\{x,y\}$-basis,

$$\chi_{\text{FM}}^{-1} = \begin{pmatrix} \frac{-J+K_x}{M^2} + \frac{i\alpha\omega}{\gamma M} & i\frac{\omega}{\gamma M} & J & 0 \\ -i\frac{\omega}{\gamma M} & \frac{-J+K_y}{M^2} + \frac{i\alpha\omega}{\gamma M} & 0 & J \\ J & 0 & \frac{-J+K_x}{M^2} + \frac{i\alpha\omega}{\gamma M} & i\frac{\omega}{\gamma M} \\ 0 & J & -i\frac{\omega}{\gamma M} & \frac{-J+K_y}{M^2} + \frac{i\alpha\omega}{\gamma M} \end{pmatrix} \tag{S.12}$$

with the poles given by $\det(\chi^{-1}(\omega)) = 0$ yielding,

$$\omega_{\text{ac}}^{\pm} = \frac{\gamma}{2M(1+\alpha^2\gamma^2)} \left[ i\alpha\gamma(K_x + K_y) \pm \sqrt{4K_xK_y - \alpha^2\gamma^2(K_x - K_y)^2} \right]$$

$$\omega_{\text{op}}^{\pm} = \frac{\gamma}{2M(1+\alpha^2\gamma^2)} \left[ i\alpha\gamma(-4J + K_x + K_y) \pm \sqrt{(4J - (K_x + K_y))^2 - (K_x - K_y)^2 - \alpha^2\gamma^2(K_x - K_y)^2} \right] \quad . \tag{S.13}$$

For antiferromagnetic dimers the different local spin frames and the different effective magnetic field lead to

$$\chi_{\text{AFM}}^{-1} = \begin{pmatrix} \frac{J+K_x}{M^2} + \frac{i\alpha\omega}{\gamma M} & i\frac{\omega}{\gamma M} & -J & 0 \\ -i\frac{\omega}{\gamma M} & \frac{J+K_y}{M^2} + \frac{i\alpha\omega}{\gamma M} & 0 & J \\ -J & 0 & \frac{J+K_x}{M^2} + \frac{i\alpha\omega}{\gamma M} & i\frac{\omega}{\gamma M} \\ 0 & J & -i\frac{\omega}{\gamma M} & \frac{J+K_y}{M^2} + \frac{i\alpha\omega}{\gamma M} \end{pmatrix} \quad , \tag{S.14}$$

with poles at

$$\omega_{\text{op},1}^{\pm} = \frac{\gamma}{2M(1+\alpha^2\gamma^2)} \left[ i\alpha\gamma(2J + K_x + K_y) \pm \sqrt{4K_x(2J + K_y) - \alpha^2\gamma^2(2J - K_x + K_y)^2} \right]$$

$$\omega_{\text{op},2}^{\pm} = \frac{\gamma}{2M(1+\alpha^2\gamma^2)} \left[ i\alpha\gamma(2J + K_x + K_y) \pm \sqrt{4K_y(2J + K_x) - \alpha^2\gamma^2(2J + K_x - K_y)^2} \right] \quad . \tag{S.15}$$

Note that the parametrization of the discussed LLG model holds for collinear dimers (FM or AFM) with out-of-plane magnetic moments, and small deviations are expected when using the realistic magnetic ground states of the dimers presented in Supplementary Note 7:.

**Collinear trimers** Similar to the dimers we consider only on-site anisotropies and isotropic exchange interactions, which yields for the effective fields in the ferromagnetic case,

$$\mathbf{B}_i^{\text{FM}} = \frac{\mathbf{e}_z}{M} \begin{cases} -J_1 - J_2 & , \quad i = \{1,3\} \\ -2J_1 & , \quad i = 2 \end{cases} \quad , \tag{S.16}$$

giving rise to the susceptibility in the $\{x,y\}$-basis,

$$\chi_{\text{FM}}^{-1} = \begin{pmatrix} \frac{-J_1-J_2+K_x}{M^2} + \frac{i\alpha\omega}{\gamma M} & i\frac{\omega}{\gamma M} & J_1 & 0 & J_2 & 0 \\ -i\frac{\omega}{\gamma M} & \frac{-J_1-J_2+K_y}{M^2} + \frac{i\alpha\omega}{\gamma M} & 0 & J_1 & 0 & J_2 \\ J_1 & 0 & \frac{-2J_1+K_x}{M^2} + \frac{i\alpha\omega}{\gamma M} & i\frac{\omega}{\gamma M} & J_1 & 0 \\ 0 & J_1 & -i\frac{\omega}{\gamma M} & \frac{-2J_1+K_y}{M^2} + \frac{i\alpha\omega}{\gamma M} & 0 & J_1 \\ J_2 & 0 & J_1 & 0 & \frac{-J_1-J_2+K_x}{M^2} + \frac{i\alpha\omega}{\gamma M} & i\frac{\omega}{\gamma M} \\ 0 & J_2 & 0 & J_1 & -i\frac{\omega}{\gamma M} & \frac{-J_1-J_2+K_y}{M^2} + \frac{i\alpha\omega}{\gamma M} \end{pmatrix} \tag{S.17}$$

with the poles given by $\det(\chi^{-1}(\omega)) = 0$ yielding,

$$\omega_1^\pm = \frac{\gamma}{2M(1+\alpha^2\gamma^2)}\left[i\alpha\gamma(K_x+K_y) \pm \sqrt{4K_xK_y - \alpha^2\gamma^2(K_x-K_y)^2}\right]$$

$$\omega_2^\pm = \frac{\gamma}{2M(1+\alpha^2\gamma^2)}\left[i\alpha\gamma(-6J_1+K_x+K_y) \pm \sqrt{(6J_1-(K_x+K_y))^2 - (K_x-K_y)^2 - \alpha^2\gamma^2(K_x-K_y)^2}\right]$$

$$\omega_3^\pm = \frac{\gamma}{2M(1+\alpha^2\gamma^2)}\bigg[i\alpha\gamma(-2J_1-4J_2+K_x+K_y)$$
$$\pm \sqrt{(2J_1+4J_2)^2 + 4K_xK_y - (4J_1+8J_2)(K_x+K_y) - \alpha^2\gamma^2(K_x-K_y)^2}\bigg] \quad . \tag{S.18}$$

For the antiferromagnetic trimer (central atom pointing along $\mathbf{e}_2 = -\mathbf{e}_z$) the effective magnetic field is given by

$$\mathbf{B}_i^{\text{AFM}} = \frac{\mathbf{e}_z}{M}\begin{cases} J_1 - J_2 &, \quad i = \{1,3\} \\ 2J_1 &, \quad i = 2 \end{cases} \tag{S.19}$$

leading to the susceptibility represented in the $\{x,y\}$-basis by

$$\chi_{\text{AFM}}^{-1} = \begin{pmatrix} \frac{J_1-J_2+K_x}{M^2}+i\alpha\omega & i\frac{\omega}{\gamma M} & -J_1 & 0 & J_2 & 0 \\ -i\frac{\omega}{\gamma M} & \frac{J_1-J_2+K_y}{M^2}+i\alpha\omega & 0 & J_1 & 0 & J_2 \\ -J_1 & 0 & \frac{2J_1+K_x}{M^2}+i\alpha\omega & i\frac{\omega}{\gamma M} & -J_1 & 0 \\ 0 & J_1 & -i\frac{\omega}{\gamma M} & \frac{2J_1+K_y}{M^2}+i\alpha\omega & 0 & J_1 \\ J_2 & 0 & -J_1 & 0 & \frac{J_1-J_2+K_x}{M^2}+i\alpha\omega & i\frac{\omega}{\gamma M} \\ 0 & J_2 & 0 & J_1 & -i\frac{\omega}{\gamma M} & \frac{J_1-J_2+K_y}{M^2}+i\alpha\omega \end{pmatrix}$$
$$\tag{S.20}$$

The poles cannot be brought into a neat form as shown for the FM case in eq. (S.18), but we can discuss the limit of an uniaxial anisotropy with $K_y = K_x$, which yields for the eigenfrequencies,

$$\omega_{\text{ac}}^\pm = \pm\frac{1}{2M}\frac{\sqrt{-\gamma^2(J_1^2(3\alpha\gamma+i)^2 + 4iJ_1K_x(\alpha\gamma+3i) - 4K_x^2)} + i\alpha\gamma^2(3J_1+2K_x) - \gamma J_1}{\alpha^2\gamma^2+1} \tag{S.21}$$

$$\omega_{\text{op},1}^\pm = \frac{1}{M}\frac{i\gamma(J_1-2J_2+K_x)}{\alpha\gamma \pm i} \tag{S.22}$$

$$\omega_{\text{op},2}^\pm = \frac{1}{2M}\frac{\pm\sqrt{-\gamma^2(J_1^2(-3\alpha\gamma+i)^2 + 4J_1K_x(-3-i\alpha\gamma) - 4K_x^2)} + i\alpha\gamma^2(3J_1+2K_x) + \gamma J_1}{\alpha^2\gamma^2+1} \quad . \tag{S.23}$$

The eigenfrequencies $\omega_{\text{op},1}$ and $\omega_{\text{op},2}$ are at high energies mainly driven by the exchange $J_1$, while $\omega_2$ yields a modes which is mainly driven by the on-site anisotropy $K_x$. To get more insights in the latter one, we can expand $\omega_{\text{ac}}$ further using the limits $\alpha \to 0$ and $J_1 \gg K_x$ yielding, $\omega_2^\pm = \pm 3\gamma K_x$, which is three times larger than the acoustical mode of the ferromagnetic trimer, eq. (S.18), and the corresponding mode of the isolated atom, eq. (S.10). Solving the generalized eigenvalue problem, $\chi^{-1}(\omega)\delta\mathbf{M} = 0$, yields the nature of this mode, which is in the global spin frame given by

$$\delta\mathbf{M}_{\text{ac}}^\pm = (\pm i, 1, \mp i, -1, \pm i, 1) \quad , \tag{S.24}$$

describing the collinear motion of the three anti-ferromagnetically coupled trimer atoms.

Note that this behaviour is fundamentally different from the AFM dimer, which only exhibits optical modes at energies related to the magnetic exchange $J_1$, see eq. (S.15). The main reason for this distinct response is that the AFM trimer has a net magnetic moment which can respond to an external magnetic field. This net magnetic moment, however, is similar to the one of the isolated atoms magnetic moment resulting in a moment per atom reduced by a factor of 3, which in turn gives rise to the enhanced excitation energy of the acoustical mode.

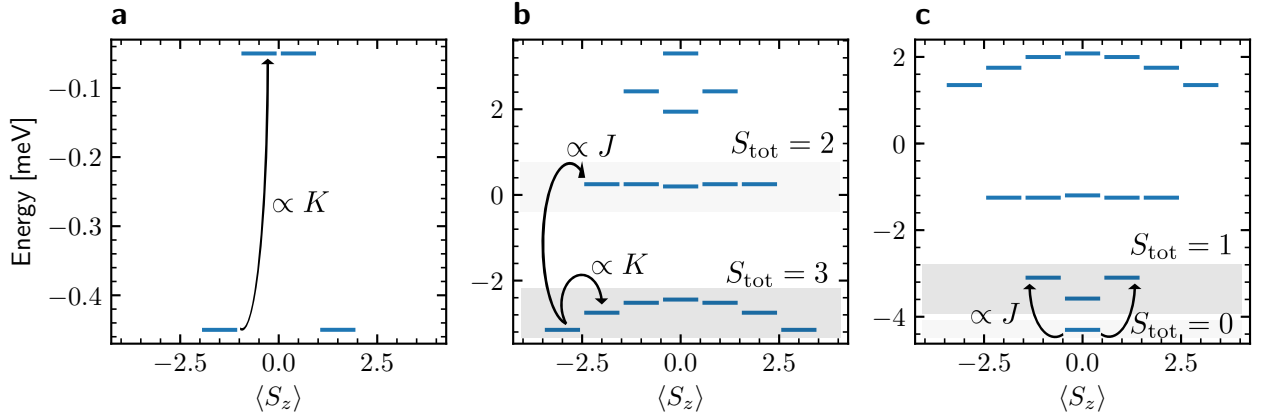

**Figure S7: Level diagrams of an isolated 3/2-spin and two magnetically coupled 3/2-spins. a** Isolated spin with a uniaxial anisotropy, $K = 0.2\,\text{meV}$. **b** Ferromagnetically coupled spins with $J = -1\,\text{meV}$ and $K_1 = K_2 = 0.2\,\text{meV}$. **c** Antiferromagnetically coupled spins with $J = 1\,\text{meV}$ and $K_1 = K_2 = 0.2\,\text{meV}$.

## Supplementary Note 9: Excitions in the quantum spin picture

Instead of the semi-classical picture which we employed in this work, magnetic atoms on surfaces are often described using the quantum spin picture. Note that on a metal due to strong hybridizations of the spin states with the surface, the ideal quantum picture of sharp isolated levels is not valid, but broadenings would have to be included. Here, we want to give an intuitive argument for the absence of an acoustical mode in an antiferromagnetically coupled dimer. Supplementary Figure **S7**a shows the level diagram of an 3/2-spin with uniaxial anisotropy, $H = K\sqrt{S(S+1)}\,\hat{S}_z^2$. This quantum spin system can be excited from the $m = -3/2$ state to the $m = -1/2$ state with a transition energy proportional to the anisotropy. The case of a ferromagnetically coupled dimer composed out of two 3/2-spins which are coupled by an isotropic magnetic exchange with $J \gg K$ is shown in Supplementary Figure **S7**b. The acoustical mode corresponds to an excitation within the subspace $S_\text{tot} = 3$, which is proportional to the magnetic anisotropy. The optical mode, however, corresponds to an excitation from the $S_\text{tot} = 3$ to the $S_\text{tot} = 2$ subspace, which is proportional to the magnetic exchange $J$. In contrast, the antiferromagnetically coupled dimer, which is shown in Supplementary Figure **S7**c, does not exhibit an accoustical mode since its ground state is the $S_\text{tot} = 0$ subspace. Thus, only the optical mode is present.

# Supplementary Note 10: Comparison of raw data for substrate, single atoms, and dimer

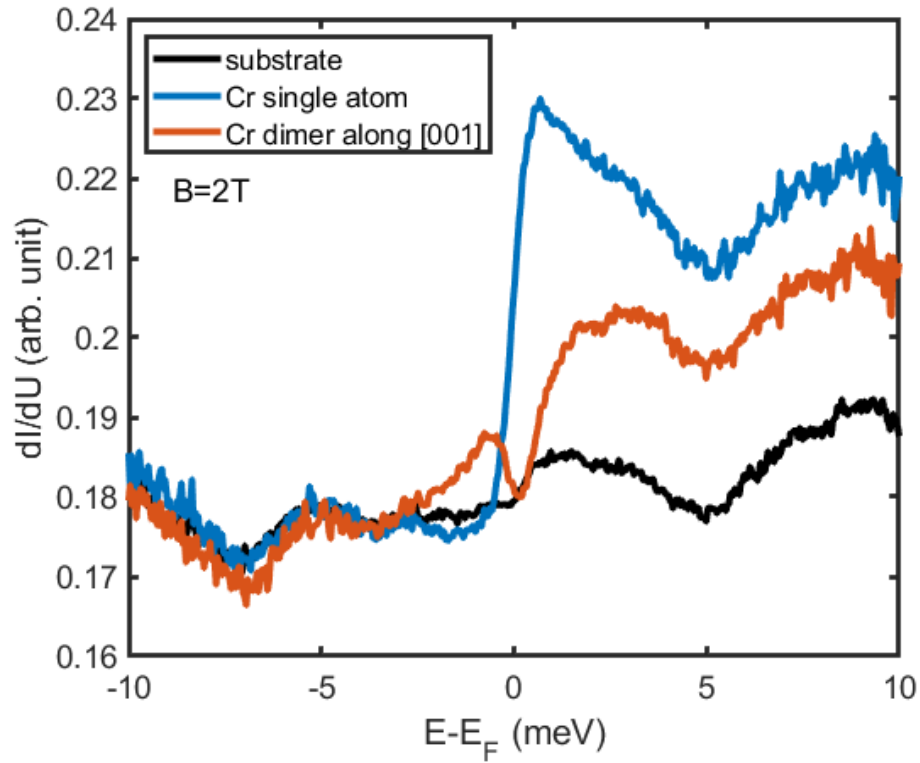

**Figure S8: Comparison of raw dI/dU signal acquired on the substrate (black line), a single Cr adatom (blue line), and a dimer aligned along the crystallographic direction [001](red line).** All measurements have been acquired using the very same W microtip under a magnetic field of $B = 2$ T applied perpendicular to the sample surface. A dip emerges close to Fermi only for the dimer, an observation which allows to link its origin to the coupling between the adatoms.

# Supplementary Note 11: Spin-resolved data acquired using functionalized superconducting tips

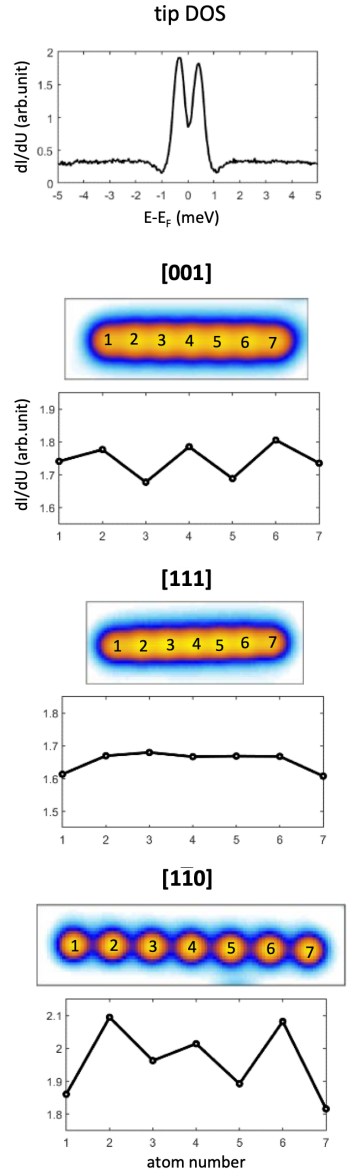

**Figure S9: Spin contrast along distinct crystallographic directions visualized using functionalized superconducting tips.** An alternating constrast between subsequent adatoms indicative of their AFM coupling is visible along [001] and [1$\bar{1}$0]. A constant amplitude indicative of FM coupling is visible along the [1$\bar{1}$1] direction. The adatoms at the end of the chains show a distinct signal because of their different local environment. All experimental observations are in line with the magnetic ground state predicted by ab-initio calculations.

# Supplementary Note 12: Spectroscopy on progressively longer chains

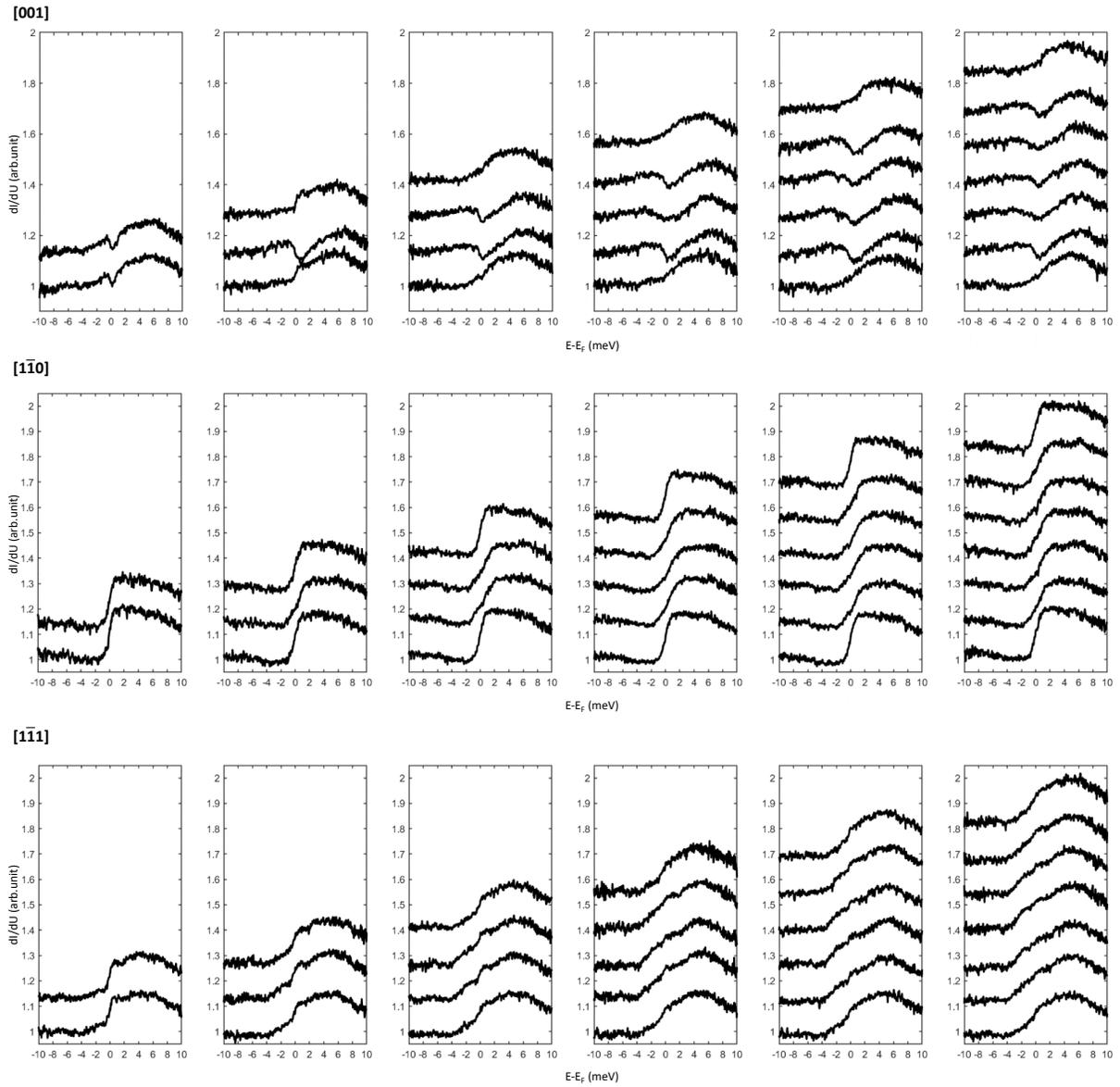

**Figure S10: Scanning tunneling spectroscopy on progressively longer chains up to 7 adatoms along the three distinct crystallographic directions scrutinized in the present study.** All measurements have been normalized to the substrate and acquired using the very same W microtip under a magnetic field of $B = 2$ T applied perpendicular to the sample surface.



\end{document}